\documentclass[12pt,a4paper]{article}
\usepackage[utf8]{inputenc}

\usepackage{graphicx}
\usepackage{amsmath}    
\usepackage{natbib}     
\usepackage[cdot,thickqspace,squaren]{SIunits} 
\usepackage{multirow}
\usepackage{array} 
\usepackage[toc,page]{appendix} 

\usepackage[english]{babel}
\usepackage{array, amsmath}
\usepackage{tikz} 
\usepackage{amssymb}
\usepackage{lineno,hyperref}

\def\matri#1{{\mbox{\boldmath  {$ #1$}\,}}}   
\def\tent#1{\oalign{\mbox{\boldmath $ #1$}\crcr\hidewidth$\scriptscriptstyle\sim$\hidewidth}}
\def\tenf#1{\oalign{\mbox{\boldmath $ #1$}\crcr\hidewidth$\scriptscriptstyle\sim$\hidewidth\crcr\hidewidth$\scriptscriptstyle\sim$\hidewidth}}
\def\vect#1{{\mbox{\boldmath  {\underline{$ #1$}}\,}}}

\title{Multiscale modeling of the elastic behavior of architectured and nanostructured Cu-Nb composite wires 
\footnote{This paper has been published on 6 May 2017 in ``International Journal of Solids and Structures'', please cite it as \cite{T_GU_elas}.}
}
\author{T. Gu, O. Castelnau, S. Forest, E. Herv\'{e}-Luanco,\\
F. Lecouturier, H. Proudhon, L. Thilly\\
}

\begin{document}
\maketitle

\section*{Abstract}

Nanostructured and architectured copper niobium composite wires are excellent candidates for the generation of intense pulsed magnetic fields ($>$90$\tesla$) as they combine both high strength and high electrical conductivity.
Multi-scaled Cu-Nb wires are fabricated by accumulative drawing and bundling (a severe plastic deformation technique), leading to a multiscale, architectured, and nanostructured microstructure exhibiting a strong fiber crystallographic texture and elongated grain shapes along the wire axis.
This paper presents a comprehensive study of the effective elastic behavior of this composite material by three multi-scale models accounting for different microstructural contents:
two mean-field models and a full-field finite element model.
As the specimens exhibit many characteristic scales, several scale transition steps are carried out iteratively from the grain scale to the macro-scale.
The general agreement among the model responses allows suggesting the best strategy to estimate the effective behavior of Cu-Nb wires and save computational time.
The importance of crystallographical and morphological textures in various cases is discussed.
Finally, the models are validated by available experimental data with a good agreement.

Keywords: Multiscale modeling, Architectured material, Polycrystalline material, Nanostructure, Elasticity, Homogenization scheme, Finite element modeling, Copper niobium composite


\section{Introduction}

\begin{figure}[!htbp]
\centering
\includegraphics[width=0.8\textwidth]{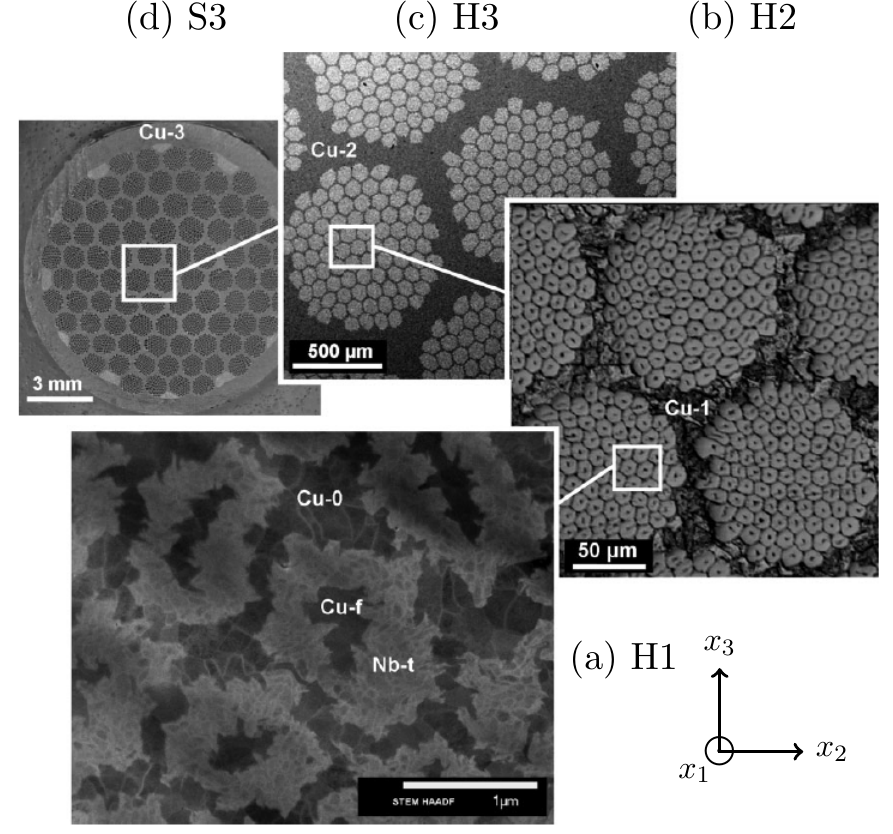}
\caption{Successive section views of the Cu/Nb/Cu nano-composite wires \citep{Dubois_2010_thesis,dubois2012cu}.
(a) Scanning Transmission Electron Microscope image showing details of the Cu/Nb/Cu elementary long fiber sections; (b), (c) and (d) Scanning Electron Microscope images showing intermediate scales and a macroscopic view of the conductor.
\textcolor{black}{In (a-c), Cu appears in dark gray and Nb in light gray; in (d), it is the reverse, due to the low magnification.} 
The diameter of specimen (d) is reduced to $0.506\,\milli\meter$ in this work by supplementary cold-drawing.
\textcolor{black}{The smaller scale for pure polycrystalline Cu or Nb (i.e. scale H0) is not shown in this figure. } 
See Section \ref{sec:intro_scale_conven} for scale conventions and their notations.
}
\label{fig:Convention_multi_scales}
\end{figure}

In recent years, there has been an increasing demand for next-generation structural nano-materials that exhibit extraordinarily high strength, electrical conductivity, hardness, ductility, thermal stability and radiation damage tolerance.
Two types of filamented and multilayered nano-composites composed of copper and niobium (i.e. Cu-Nb nano-composite wires and laminates) are highlighted among them \citep{misra2010structural}.
These two Cu-Nb nano-composites are fabricated respectively by two different severe plastic deformation techniques:
Accumulative Drawing and Bundling \citep{dupouy1996microstructural,thilly2002high}
and Accumulative Roll Bonding \citep{lim2009length,beyerlein2014emergence}.

A series of Cu-Nb nano-composite wires is illustrated in Fig. \ref{fig:Convention_multi_scales}, referred to as co-cylindrical structure in \cite{Dubois_2010_thesis}: a multiscale Cu matrix embedding parallel Nb nano-tubes filled with Cu nano-filaments.
These nano-composite conductors are excellent candidates for generation of intense pulsed magnetic fields ($>$90\tesla).
According to \citep{spencer2004established,buard2013special,halperin2013high,frydman2014high}, intense magnetic fields are becoming essential experimental and industrial tools.
To generate them, the conductors for the winding coils must combine both high mechanical strength and high electrical conductivity.
In \citep{thesis3,vidal2007cu}, a conductor presenting an Ultimate Tensile Strength as large as $1.9\,\giga\pascal$ at $77\kelvin$ is obtained together with an electrical conductivity of $1.72\,\reciprocal{\micro \ohm} \reciprocal{\centi\meter}$.

To predict the behavior of such a composite, the main challenge remains in the understanding of the complex interaction between the different material phases, and the architecture, in particular when the Cu-Nb composite is fabricated by severe plastic deformations where the elementary physical deformation mechanisms are modified by grain sizes.
In this field, combining material characterization and multi-scale modeling is mandatory.
The previous studies on the Cu-Nb nano-composite wires and laminates deal with
deformation mechanisms \citep{thilly2001deformation},
hardness \citep{thilly2002size},
Cu-Nb interfaces \citep{beyerlein2014emergence,carpenter2015interface},
Ultimate Tensile Strength \citep{vidal2007cu},
plastic flow stability \citep{misra2007plastic},
Bauschinger effect \citep{thilly2007evidence,zhang2013role,badinier2014bauschinger},
yield criterion \citep{thilly2009new},
texture evolution \citep{lim2009length,lee2012heterophase,hansen2013modeling},
thermal stability and internal stresses \citep{dubois2010thermal,dubois2012cu},
and electrical conductivity \citep{thesis3,Dubois_2010_thesis,gu2015modelisation}.

The present work concentrates on the multiscale modeling of the \emph{anisotropic elastic behavior} of architectured and nanostructured Cu-Nb composite \emph{wires}.
Three multi-scale methods will be introduced:
two mean-field homogenization models (Standard and Generalized Self-Consistent schemes, denoted respectively SSC and GSC hereafter, see Table \ref{tab:abbreviations} for the abbreviations) and a full-field Finite Element Method (FEM) with periodic boundary conditions (denoted PH, for Periodic Homogenization).
These models essentially differ by the microstructural information they are based on for the estimation of the effective behavior.
Here, the SSC scheme will be used to describe the elastic response of polycrystals made of Cu or Nb grains, but also for the estimation of a random mixture of Cu and Nb phases.
The GSC scheme takes into account a specific filament/nanotube Cu/Nb/Cu microstructure.
PH assumes a periodic microstructure, and its response will be compared with the ones obtained by SSC and GSC approaches.

\begin{table}[!htbp]
\centering
\begin{tabular}{|c|c|}
\hline
SSC & Standard Self-Consistent\\
\hline
GSC & Generalized Self-Consistent\\
\hline
FEM & Finite Element Method\\
\hline
PH & Periodic Homogenization\\
\hline
HEM & Homogeneous Equivalent Medium\\
\hline
RVE & Representative Volume Element\\
\hline
\end{tabular}
\caption{The abbreviations used in this work.}
\label{tab:abbreviations}
\end{table}

The SSC scheme is known as a homogenization theory well adapted to estimate the mechanical behavior of polycrystals.
This mean-field homogenization method is based on a statistical description of the microstructure of  polycrystalline aggregates.
The underlying microstructure, described by \cite{kroner1978self}, corresponds to perfect disorder with infinite graduation of size.
The development of the SSC model for heterogeneous elasticity goes back to \citep{hill1965self,budiansky1965elastic,Kneer_1965,willis1977bounds,kroner1978self}.
Later on, the model has been extended to  visco-plastic, elasto-plastic, and elasto-visco-plastic properties, e.g. see \citep{molinari1987self,castaneda1998nonlinear,lebensohn2011full,yoshida2011micromechanical,Vu-et-al-MSMSE-2012}.
Analysis of the intraphase stress and strain heterogeneity obtained by the SSC scheme, and its comparison with full-field reference calculations, can be found e.g. in
\citep{castaneda1998nonlinear,brenner2004mechanical,lebensohn2011full}.

The GSC scheme is another mean-field homogenization method taking into account particular morphologies where multi-layered fibers are considered.
This kind of morphology has been first studied by \cite{hashin1962elastic} who has developed variational bounding methods applied to a Composite Sphere Assembly made of an arrangement of homothetic two-layers spheres.
Then, \cite{hashin1964elastic} have applied these variational bounding methods to exhibit bounds for the five independent elastic moduli of the two-dimensional analogue of the Composite Spheres Assembly.
\cite{christensen1979solutions} have then derived an estimation for the elastic behavior of such sphere or fiber-reinforced composite by considering a representative two-layers concentric sphere/cylinder embedded in a fictitious homogeneous medium representing the Homogeneous Equivalent Medium (HEM).
Their method is known as the ``three-phase model".
\cite{herve1993n,herve1995elastic} have then extended \cite{christensen1979solutions}'s approach to multi-coated sphere or fiber-reinforced composite thanks to the ``($n+1$)-phase model''.
This model is used here to study the elastic behavior of the Composite Cylinders Assembly present at different scales (Fig. \ref{fig:Convention_multi_scales}).
The GSC scheme has also been extended to visco-elastic behavior \citep{beurthey2000structural}, nonlinear behavior \citep{zaoui1997structural}, diffusion \citep{care2004application}, thermal conductivity \citep{herve2002thermal}.
Interphase effects have also been taken into account in \cite{hashin2002thin} and \cite{herve2014elastic}.
Application to the diffusion phenomena is presented in \cite{gu2015modelisation,herve2016multiscale,joannes2016multiscale}.

With the increase in the computational performance and the number of available numerical software products, computational full-field homogenization methods have gained attention.
Unlike mean-field approaches (e.g., SSC and GSC schemes), the full-field methods (e.g. based on FEM) applied to Representative Volume Element (RVE) can describe the detailed experimental microstructure and provide the complex stress/strain fields inside the different phases at the expense of increased computational time.
Some full-field methods for polycrystalline aggregates were developed by \cite{ghosh1995multiple,ghosh1996two} in terms of a special class of finite element based on Vorono\"i cells.
Making use of such full-field homogenization models, several linear material behaviors were analyzed:
the effective thermal conductivity \citep{flaquer2007effect},
the effective thermoelastic properties and residual stresses \citep{wippler2011homogenization},
and the effective elastic properties with a statistical description \citep{kanit2003determination,fritzen2009periodic,bohlke2010elastic}.
More complex non-linear mechanical behaviors are also studied for Face-Centered Cubic and Body-Centered Cubic polycrystalline aggregates in many aspects, such as in \citep{cailletaud2003some,bohlke2009numerical,schneider2010plastic,fritzen2011nonuniform,klusemann2012investigation}.
For taking the anisotropic morphological textures into account, three-dimensional Vorono\"i mesh generation techniques have been proposed \citep{barbe2001intergranular,fritzen2009periodic,fritzen2011nonuniform}.
In addition, \citep{yaguchi2005accuracy,fritzen2009periodic,bohlke2010elastic} have also compared full-field model results with the previously mentioned mean-field ones for polycrystalline aggregates.

The full-field FEM can also consider a specific periodic microstructure, such as Composite Spheres Assembly in \citep{llorca2000elastic,michel2009nonuniform} and Composite Cylinders Assembly in \cite{gu2015modelisation}.
In this homogenization method, the microstructure unit cell is subjected to periodic boundary conditions \citep{besson2009non}, therefore it is named as PH model in this work.
\cite{llorca2000elastic} compared the three above mentioned models SSC, GSC, and PH for isotropic elastic behaviors of the Composite Spheres Assembly architecture.

Despite a wealth of literature works on polycrystalline elasticity, we have found that the following three points are still missing:
1. No systematic analysis of the simultaneous contributions of morphological and crystallographic textures to the anisotropic elastic properties for Cu and Nb, especially for the specific crystallographic textures encountered in Cu-Nb wires.
2. No application of current homogenization methods to the complex architectures (i.e. Composite Cylinders Assembly) of recent Cu-Nb composites.
\textcolor{black}{3. No experimental comparison with in-situ X-ray/neutron diffraction data \citep{thilly2006plasticity,thilly2007evidence,thilly2009new} for Cu-Nb wires.} 

Therefore, the objectives of this paper are threefold:
1. provide a homogenization model for Cu polycrystals and Nb polycrystals taking the crystallographic and morphological textures into account;
2. provide a multi-scale homogenization procedure to model the architectured and nanostructured Cu-Nb composite wires;
\textcolor{black}{3. provide a quantitative comparison with elastic strain measured by X-ray/neutron diffraction during uniaxial loading tests.} 
The outline of the article is as follows.
The architecture and nano-structure of Cu-Nb composite wires are described in Section \ref{sec:material_description}.
In order to reproduce the effective elastic behavior of this material, three multi-scale methods, i.e. SSC, GSC and PH are presented in Section \ref{sec:models}.
In Section \ref{sec:H0}, the Cu polycrystals and the Nb polycrystals are considered separately.
Then in Section \ref{sec:H1_to_H3}, several scale transitions of architectured Cu-Nb composites are performed to determine the effective elastic behavior of Cu-Nb wires up to macro-scale.
In Section \ref{sec:discussions}, the importance of micro-parameters and the best modeling strategies are discussed for various textures and material properties.
Finally, the model results are validated by comparison with available experimental data.

Throughout this work, the following notation is used:
$x$ for scalars, $\vect{x}$ for vectors, $\tent{x}$ for 2nd-order tensors, $\tenf{x}$ for 4th-order tensors, $\cdot$ for single contraction, $:$ for double contraction, $\otimes$ for tensor product, $\widetilde{x}$ for effective (or homogenized) property and $\bar{x}=\langle x \rangle$ for volume average.


\section{Material description}
\label{sec:material_description}

\subsection{Fabrication process}

Cu-Nb nano-composite wires are fabricated via a severe plastic deformation process, based on Accumulative Drawing and Bundling (series of hot extrusion, cold drawing and bundling stages) according to \citep{dupouy1996microstructural,thilly2002size,thilly2002high,Dubois_2010_thesis}:
a Cu wire is inserted into a Nb tube, itself inserted into a Cu tube. The structure is extruded and drawn, then cut into $85$ smaller pieces with hexagonal cross section.
These pieces are then bundled and inserted into a new Cu tube.
The new composite structure is again extruded and drawn.
And so on.
In the present work, Accumulative Drawing and Bundling is repeated three times, leading to copper based architectured and nanostructured composite wires (so-called ``co-cylindrical Cu/Nb/Cu'' wires) which are composed of a multi-scale Cu matrix embedding $85^3$ Nb nanotubes containing $85^3$ Cu nanofibers, as illustrated in Fig. \ref{fig:Convention_multi_scales}.
The used Cu is Oxygen-Free High Conductivity (OFHC).
It is noted that, unlike Cu, Nb tubes are introduced only at the very first fabrication stage.
Therefore, Nb nanotubes (denoted Nb-t in Fig. \ref{fig:Convention_multi_scales}(a)) are all deformed together during the iterative Accumulative Drawing and Bundling, and they exhibit the same microstructure and similar characteristic sizes.
The Cu-f and Cu-0 regions (Fig. \ref{fig:Convention_multi_scales}(a)) are introduced at the beginning of the process, while the Cu-1, Cu-2, Cu-3 (Fig. \ref{fig:Convention_multi_scales}(b-d)) are introduced successively during the three steps of Accumulative Drawing and Bundling;
different microstructure are thus expected for the different Cu-i regions ($i=f$, 0, 1, 2 and 3).


\subsection{Scales}
\label{sec:intro_scale_conven}

For a wire with a final diameter\footnote{Following \citep{vidal2007cu, thilly2009new}, all dimensions are given in the $x_2$-$x_3$ cross-section, i.e. perpendicular to the wire axis $x_1$, see Fig. \ref{fig:Convention_multi_scales} for the coordinate system.}
of 0.506 $\milli\meter$, Nb nanotubes (average wall thickness $\delta_{\text{Nb-t}}$=88 $\nano\meter$ and total volume fraction $X_{\text{Nb-t}}$=20.8\%) are filled with Cu-f copper filaments (diameter $\delta_{\text{Cu-f}}$=130 $\nano\meter$ and volume fraction $X_{\text{Cu-f}}$=4.5\%), separated by the finest Cu-0 copper channels (width $\delta_{\text{Cu-0}}$=93 $\nano\meter$ and volume fraction $X_{\text{Cu-0}}$=17.7\%);
groups of 85 Cu/Nb/Cu elementary long fibers are separated by Cu-1 copper channels (width $\delta_{\text{Cu-1}}$=360 $\nano\meter$ and $X_{\text{Cu-1}}$=9.6\%).
The width of Cu-2 copper channel is $\delta_{\text{Cu-2}}$=3.9 $\micro\meter$ ($X_{\text{Cu-2}}$=19.9\%).
Finally, the group of $85^3$ elementary patterns is embedded in an external Cu-3 copper jacket ($\delta_{\text{Cu-3}}$=21.1 $\micro\meter$ and $X_{\text{Cu-3}}$=27.5\%).

For multiscale modeling of the effective elastic behavior of these Cu-Nb wires, the following scale conventions will be used:
(1) Homogenization at the highest magnification scale, looking directly at each polycrystalline Cu or Nb phase, is labeled as H0 (Homogenization 0);
(2) Then, homogenization of the assembly of $85^1$ elementary Cu-f/Nb-t/Cu-0 long fibers is labeled as H1;
(3) Iterative homogenization of the effective Cu-Nb composite zone of H($n-1$) embedded in the Cu-($n-1$) matrix (with $n=2$ or 3), is labeled as H$n$, i.e. H$n$ provides the effective behavior of an assembly of $85^{n}$ elementary patterns.
The effective stiffness tensors denoted $(\widetilde{\tenf{C}})_\text{H0}$ at the scale H0 of polycrystalline aggregates (i.e. Cu and Nb polycrystals), and those denoted $(\widetilde{\tenf{C}})_\text{Hi}$ ($i=1$, 2, 3) at scales H$i$ of assembly of $85^\text{i}$ elementary long fibers, will be obtained from the homogenization models.

Finally, the scale S3 (see Fig. \ref{fig:Convention_multi_scales}(d)) is defined here as a single cylinder-shaped structure with two layers: effective Cu-Nb composite zone of H3 (containing $85^3$ elementary patterns) surrounded by the external Cu-3 jacket.
The structural problem S3 will be solved by FEM to compute the stiffness $(\widetilde{\tenf{C}})_\text{S3}$.
Then, this effective stiffness will be compared with the available experimental data.


\subsection{Morphological and crystallographic textures}
\label{sec:intro_textures}

The microstructural state of Cu-Nb wires has been studied by Scanning Electron Microscope, Scanning Transmission Electron Microscope and X-ray diffraction by \citep{vidal2007cu,thilly2009new,dubois2010thermal,Dubois_2010_thesis,dubois2012cu}.
In the present work, we define $\ell$ as the average grain length along the longitudinal wire axis $x_1$ and $d$ as the average grain diameter in the transverse $x_2$-$x_3$ cross-section.
The previous studies have shown that the morphological texture exhibits highly elongated grains along $x_1$ because of iterative severe plastic extrusion and drawing, therefore $\ell \gg d$ for Cu and Nb phases.

Because of the multi-scale structure, different types of copper matrix channels are present in the matrix:
(i) Cu channels with a width $\delta_{\text{Cu-i}}$ (i=2,3) larger than a few micrometers (so-called ``large'' Cu channels) are mainly composed of grains with a transverse size $d$=200-400 $\nano\meter$ (a typical microstructure of cold-worked material); (ii) On the other hand $\delta_{\text{Cu-i}}$ (i=f,0,1) lies in the sub-micrometer range (so-called ``fine'' Cu channels), with only a few grains located between the Cu-Nb interfaces: in this case, grain width $d_{\text{Cu-i}}$ varies from $\delta_{\text{Cu-i}} /3$ to $\delta_{\text{Cu-i}}$.
In addition, the grain size of Nb tubes $d_{\text{Nb-t}}$ is comparable with the tube width ($d_{\text{Nb-t}} \approx \delta_{\text{Nb-t}}$).

The overall crystallographic texture of a Cu-Nb co-cylindrical composite sample at a diameter of 3.5 $\milli\meter$ has been estimated by X-ray diffraction in \cite{Dubois_2010_thesis}.
The specimens with smaller diameter 0.506 $\milli\meter$ considered in this work are believed to display very similar crystallographic texture, as confirmed by preliminary EBSD results.
X-ray diffraction has shown that Cu phases exhibit strong $\langle111\rangle$ fiber texture with the remnant $\langle100\rangle$ fiber, while a single-component $\langle110\rangle$ fiber texture is observed in Nb phases.
Due to extrusion and drawing along the wire direction ($x_1$) in the fabrication process, $x_1$ is also the symmetry axis of these fibers.
In Cu, the volume fractions of $\langle100\rangle$ and $\langle111\rangle$ components are found to be 37\% and 58\% respectively (with 5\% of an additional random component).
In Nb phase, the volume fraction of $\langle110\rangle$ fiber is 99\% (with 1\% random component).
The associated texture spread (Full Width at Half Maximum) of individual components are indicated in Table \ref{tab:crystallographic_textures_Cu_Nb_Dubois_2010}.

\begin{table}[!htbp]
\centering
\begin{tabular}{|c|c|c|c|c|c|}
\hline
\multicolumn{2}{|c|}{Fiber} & Volume & \multicolumn{2}{c|}{FWHM (in $deg$)} \\
\cline{4-5}
\multicolumn{2}{|c|}{textures} & fractions & ~~$\Delta \Phi$~~ & $\Delta \Phi_1$ \\
\hline
\multirow{2}{*}{Cu} & $\langle100\rangle$ & 37\% & 10.1 & 11.5 \\
\cline{2-5}
 & $\langle111\rangle$ & 58\% & 8.3 & 10.2 \\
\hline
Nb & $\langle110\rangle$ & 99\% & 6.2 & 6.8 \\
\hline
\end{tabular}
\caption{Overall crystallographic textures and corresponding Full Width at Half Maximum (FWHM) of individual components, for Cu and Nb polycrystals.
These fiber textures (symmetry axis $x_1$) of the Cu-Nb composites were determined by X-ray diffraction \citep{Dubois_2010_thesis}.
In addition, Cu and Nb contain also 5\% and 1\% random components respectively.}
\label{tab:crystallographic_textures_Cu_Nb_Dubois_2010}
\end{table}

In the fabrication process, polycrystalline Nb tubes are always deformed simultaneously, thus they all display the same crystallographic texture.
However, the Cu polycrystals are introduced successively at the three steps of Accumulative Drawing and Bundling, therefore the crystallographic textures are different for each Cu-$i$ ($i=f$, 0, 1, 2 and 3).
Local textures at the very fine scale need to be characterized by EBSD, a work currently in progress.
In the present work, for the sake of simplicity, we consider the same Cu crystallographic textures, as determined from X-ray diffraction in Table \ref{tab:crystallographic_textures_Cu_Nb_Dubois_2010}, at all scales of the Cu-Nb wires.
In other words, effective elastic behaviors of all the Cu-$i$ are assumed to be identical.
It will be shown that this approximation is sufficient to predict the elastic behavior for Cu-Nb wires based on the available mechanical data.

\begin{figure}[!htbp]
\centering
\begin{tikzpicture}
\node[above=0cm] at (0,0) {\includegraphics[width=7.5cm]{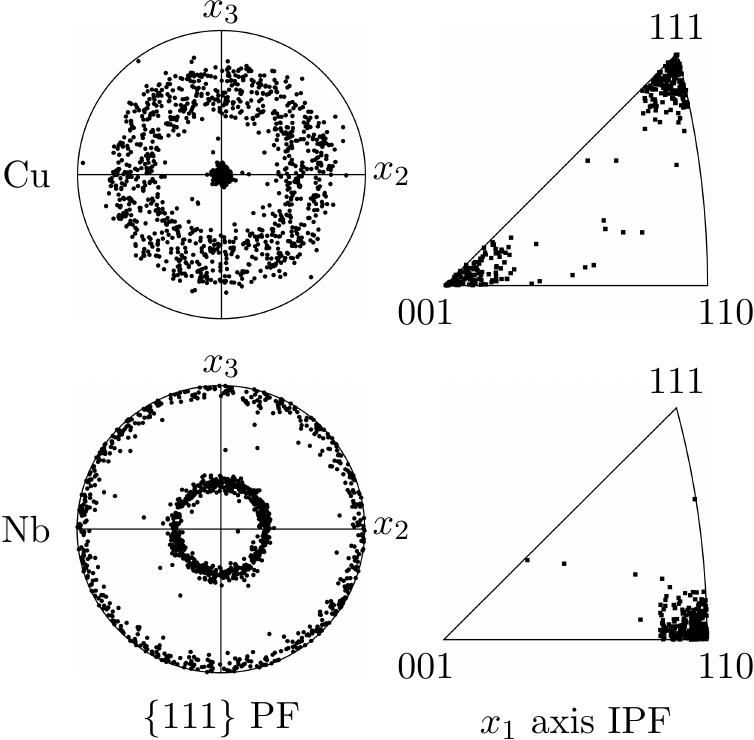}};
\end{tikzpicture}
\caption{
$\{$111$\}$ Pole Figure and $x_1$ axis Inverse Pole Figure of Cu polycrystals and Nb polycrystals.
Grain orientations were generated by LaboTex based on experimental X-ray textures. \textcolor{black}{Central stereographic projections}.
}
\label{fig:PF_IPF_Cu_Nb}
\end{figure}

From the Orientation Distribution Function described above, we used the software LaboTex \footnote{Software for crystallographic textures - http://www.labosoft.com.pl/.} to generate two sets of $40000$ discrete orientations each (one for Cu and one for Nb) that have been used to generate the microstructure in the three scale transition models.
Figure \ref{fig:PF_IPF_Cu_Nb} shows the obtained Pole Figures and Inverse Pole Figures for both Cu and Nb, using $300$ orientations randomly chosen among each of the larger sets of $40000$ orientations.
The full set of $40000$ orientations has been used within the SSC scheme to estimate the effective behavior (section \ref{sec:model_SSC}), whereas subsets of 1000 (respectively 100) orientations randomly chosen among the 40000 have been used for parallelepipedic (resp. Vorono\"i) tessellations (section \ref{sec:model_PH}).

\subsection{Anisotropic elastic properties}
\label{sec:intro_notations}

The Cu-Nb wires are made of Face-Centered Cubic Cu and Body-Centered Cubic Nb grains.
According to \citep{epstein_1965_cu,hiki_1966_cu,carroll_1965_nb}, the cubic elastic constants $C_{ij}$ (Voigt convention) of Cu and Nb single crystals, expressed in a reference frame attached to the cubic crystal lattice,
are provided in Table \ref{tab:Monocrystal_Cu_Nb}, and they will be used in H0 for predicting the effective elastic behavior of polycrystalline aggregates.
The Zener anisotropy factor $Z$ \citep{zener1948elasticity}, defined as
\begin{equation}
Z=\displaystyle{\frac{2C_{44}}{(C_{11}-C_{12})}} \;,
\label{eq:Zener}
\end{equation}
is a measure of the elastic anisotropy, having $Z$=1 for an isotropic material.
As shown in Table \ref{tab:Monocrystal_Cu_Nb}, Cu and Nb single crystals exhibit strong anisotropy.
The Young's modulus in a $\langle111\rangle$ direction is about three times higher (respectively, half smaller) than the one in a $\langle100\rangle$ direction in Cu (respectively, Nb) single crystals.

\begin{table}[!htbp]
\centering
\begin{tabular}{|c|c|c|c|c|}
\hline
Single & $C_{11}$ & $C_{12}$ & $C_{44}$ & \multirow{2}{*}{$Z$} \\
crystal & ($\giga\pascal$) & ($\giga\pascal$) & ($\giga\pascal$) &  \\
\hline
Cu & 167.20 & 120.68 & 75.65 & 3.25 \\
\hline
Nb & 245.60 & 138.70 & 29.30 & 0.55\\
\hline
\end{tabular}
\caption{Cubic elastic constants of Cu and Nb single crystals, expressed in the crystal lattice (Voigt convention).}
\label{tab:Monocrystal_Cu_Nb}
\end{table}

The components $C_{ijkl}^{(r)}$ of the single crystal elastic moduli $\tenf{C}^{(r)}$ expressed in a generic global coordinate system can be calculated from the components  $C_{mnpq}$ of the independent elastic stiffness $\tenf{C}$ in the lattice coordinate system
\begin{equation}
C_{ijkl}^{(r)}=Q_{mi}^{(r)}Q_{nj}^{(r)}Q_{pk}^{(r)}Q_{ql}^{(r)}C_{mnpq}
\label{eq:rotation_matrix}
\end{equation}
where $Q_{mi}^{(r)}$ are the components of the rotation matrix $\matri{Q}^{(r)}$ associated with the crystal orientation $r$.
The corresponding Euler angles $(\psi, \theta, \phi)^{(r)}$ are needed to determine $\matri{Q}^{(r)}$ \citep{slaughter2002linearized}.

Due to the material processing, the architecture of Cu-Nb wires, morphological and crystallographic textures are axisymmetric with respect to axis $x_1$.
As a result, the effective material behavior at scale H$i$ ($i=0$, 1, 2, 3) is expected to be transversely isotropic.
Anisotropic elasticity is then expressed by five independent constants \citep{herve1995elastic}:
longitudinal Young's modulus $\widetilde{E}_{1}$,
Poisson's ratio under longitudinal load $\widetilde{\nu}_{12}$,
longitudinal shear modulus $\widetilde{\mu}_{12}$,
transverse shear modulus $\widetilde{\mu}_{23}$,
plane-strain bulk modulus $\widetilde{K}_{23}$.
Incidentally, for transverse isotropy, transverse Young's moduli $\widetilde{E}_2 \equiv \widetilde{E}_3$.
In this work, they will be noted as $\widetilde{E}_{2,3}$ and can also be derived from the other parameters:
\begin{equation}
\widetilde{E}_{2,3}=
\frac{4\widetilde{E}_{1}\, \widetilde{K}_{23}\, \widetilde{\mu}_{23}}{4\widetilde{K}_{23}\widetilde{\mu}_{23}\widetilde{\nu}_{12}^2
+ \widetilde{E}_{1}\left( \widetilde{K}_{23} + \widetilde{\mu}_{23} \right)}
\label{eq:E_T_E_nu_K_mu}
\end{equation}
It is also worth noting that $\widetilde{C}_{11}=\widetilde{E}_{1}+4\widetilde{\nu}_{12}^{2}\widetilde{K}_{23}$,
$\widetilde{C}_{22}=\widetilde{C}_{33}=\widetilde{K}_{23}+\widetilde{\mu}_{23}$,
$\widetilde{C}_{12}=\widetilde{C}_{13}=2\widetilde{\nu}_{12}\widetilde{K}_{23}$,
$\widetilde{C}_{44}=\widetilde{\mu}_{23}$,
$\widetilde{C}_{55}=\widetilde{C}_{66}=\widetilde{\mu}_{12}$,
$\widetilde{C}_{23}=\widetilde{K}_{23}-\widetilde{\mu}_{23}$
(with $\widetilde{C}_{23}=\widetilde{C}_{22}-2\widetilde{C}_{44}$ in the case of transverse isotropy),
where $\widetilde{C}_{ij}$ denotes the components of $\widetilde{\tenf{C}}$ making use of the Voigt notation.


\section{Homogenization strategies}
\label{sec:models}

\begin{figure}[!htbp]
\centering
\begin{tikzpicture}
\node at (0,0) {\includegraphics[width=0.95\textwidth]{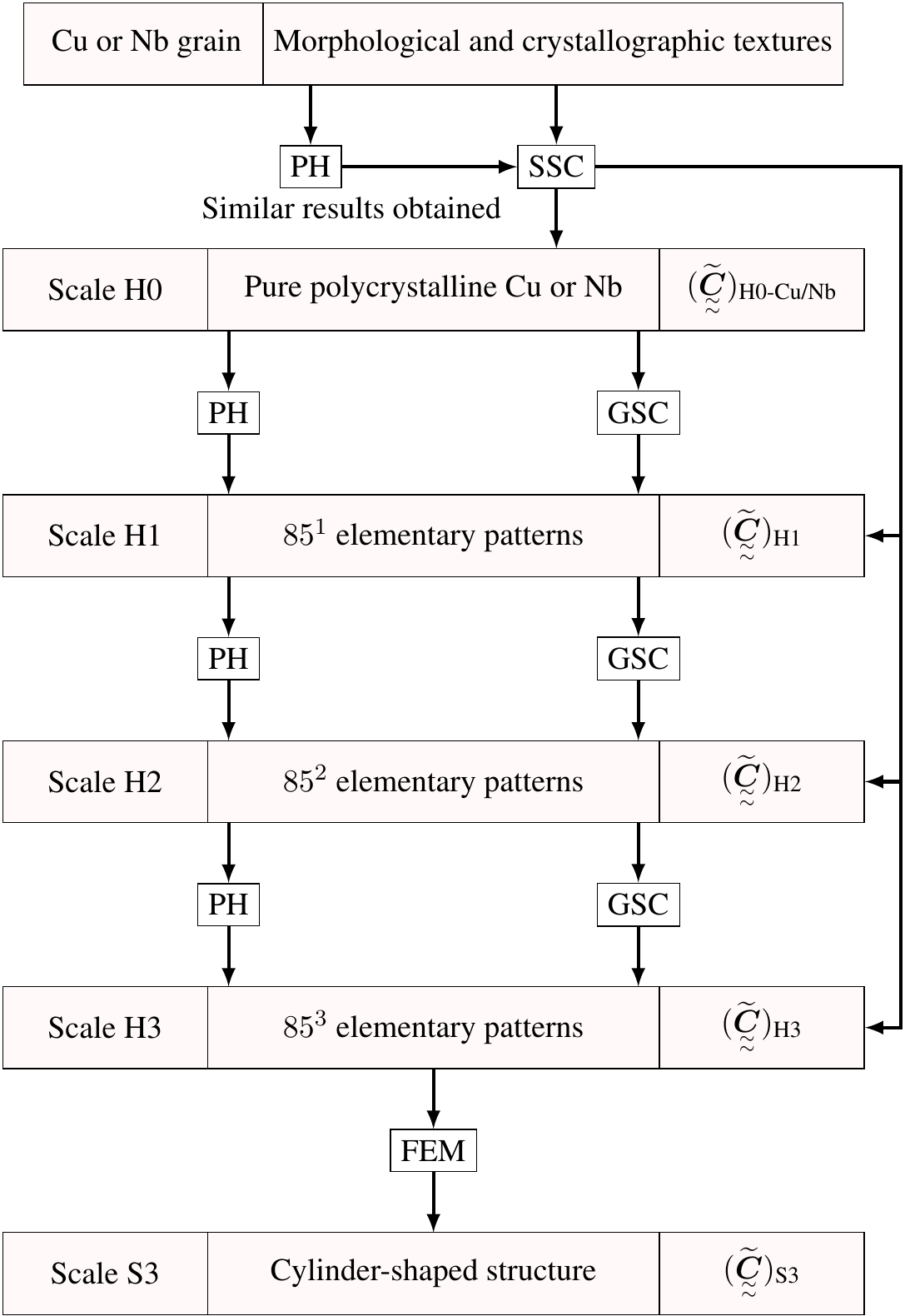}};
\end{tikzpicture}
\caption{\textcolor{black}{Overview chart of the iterative scale transition steps. The considered scales and scale transition models are mentioned, together with the obtained effective behavior.}
}
\label{fig:overview_chart}
\end{figure}

\textcolor{black}{The homogenization strategy used in this work, including the corresponding scales and model used, is illustrated in Fig. \ref{fig:overview_chart}}.

\subsection{Mean-field standard self-consistent scheme}
\label{sec:model_SSC}

\begin{figure}[!htbp]
\centering
\begin{tikzpicture}
\node at (0,0) {\includegraphics[height=5.7cm]{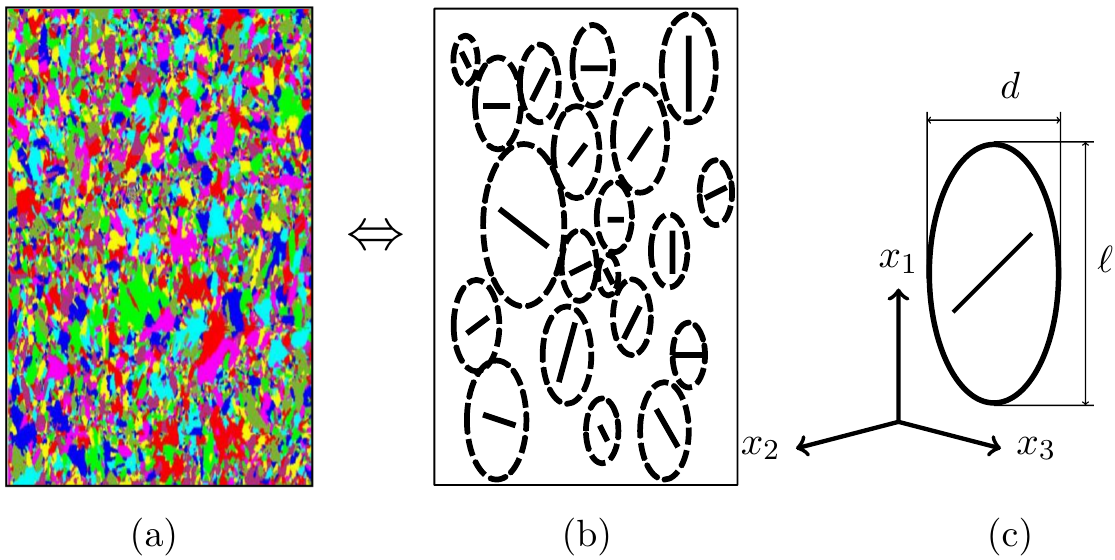}};
\end{tikzpicture}
\caption{Schematic geometry of the SSC model:
(a) Perfectly disordered mixture of the grains, being similar with the real morphological texture in polycrystals.
(b) Statistically equivalent medium composed of randomly mixed \textit{mechanical phases}.
A \textit{mechanical phase} is denoted by the set of grains having the same crystal orientation.
(c) The overall (mean) shape of all \textit{mechanical phases} is taken \textcolor{black}{spheroidal}, elongated along direction $x_1$ and with aspect ratio $\ell/ d$.}
\label{fig:SSC_Zoom0}
\end{figure}

The mean-field SSC scheme is used in the present work to estimate the effective elastic property $\widetilde{\tenf{C}}$ of individual Cu and Nb polycrystals.
It is also used to estimate the behavior of a fictitious material in which Cu and Nb grains would be randomly mixed all together, with volume fractions and textures introduced above, in order to check the impact of the particular architectured microstructure of our specimens.

Owing to the random character of the microstructure with all grains playing geometrically similar roles, the SSC scheme is especially suited for polycrystals \citep{kroner1978self,thomas2011mechanics}.
The SSC scheme relies on specific microstructure exhibiting a sufficiently irregular mixture of grains and infinite size graduation as illustrated in Fig. \ref{fig:SSC_Zoom0}(a).
As in \citep{lebensohn2011full, thomas2011mechanics}, one defines a \textit{mechanical phase} ($r$) as denoting the set of all grains in the microstructure that share the same elastic properties, that is exhibiting the same crystal orientation; those grains have however different shapes and environment.
The material can thus be statistically described as an equivalent aggregate filled with mechanical phases of different size and surrounding, and distributed randomly.
It is supposed that the phases exhibit, \emph{on average}, a \textcolor{black}{spheroidal} shape, and therefore the \emph{mean} stress and strain inside those phases can be estimated with the Eshelby inclusion problem; but note that this does not mean that stress and strain are uniform inside individual phases, see e.g. \cite{castaneda1998nonlinear,brenner2004mechanical}.
Here, spheroids corresponding to the mean grain shape are elongated along the longitudinal direction $x_1$, with length $\ell$ and width $d$, as in Fig. \ref{fig:SSC_Zoom0}(c).
The associated ``grain aspect ratio'' $\ell/d$ statistically defines the morphological texture of the polycrystalline aggregate.

For elastic polycrystals, local and effective constitutive relations read respectively:
\begin{equation}
\tent{\sigma}(\vect{x})=\tenf{C}(\vect{x}):\tent{\varepsilon}(\vect{x}),
~~~~\tent{\overline{\sigma}}=\tenf{\widetilde{C}}:\tent{\overline{\varepsilon}}
\label{eq:local_effective_constitutive_relations}
\end{equation}
with $\tenf{\widetilde{C}}$ the effective stiffness tensor
\begin{equation}
\widetilde{\tenf{C}}=<\tenf{C}(\vect{x}):\tenf{A}(\vect{x})>
\label{eq:C_eff_C_local_A}
\end{equation}
where \tenf{A}(\vect{x}) is the strain localization tensor defined as:
\begin{equation}
\tent{\varepsilon}(\vect{x})=\tenf{A}(\vect{x}):\tent{\overline{\varepsilon}}\,,
\label{eq:SSC_definition_A}
\end{equation}
and $\tent{\overline{\varepsilon}}$ being the macroscopic applied strain.
In elastic polycrystals, the local stiffness tensor is a uniform property inside grains.
The quantity $\tenf{C}(\vect{x})$ in Eq.(\ref{eq:local_effective_constitutive_relations}) can therefore be replaced by the corresponding homogeneous values $\tenf{C}^{(r)}$ of the considered \textit{mechanical phase} ($r$) defined previously.
Similar substitution can be made in Eq.(\ref{eq:C_eff_C_local_A}) leading to:
\begin{equation}
\tenf{\widetilde{C}}=\sum_{n=1}^{r}f^{(r)}\tenf{C}^{(r)}:\tenf{A}^{(r)}
\label{eq:C_eff_C_local_A_r}
\end{equation}
where $\ldotp^{(r)}$ indicates the average over the volume of phase ($r$), e.g. $\tenf{A}^{(r)}=\left \langle \tenf{A}(\vect{x}) \right \rangle_{(r)}$, and $f^{(r)}$ denotes volume fraction of phase ($r$).

To estimate the phase-average stress and strain, phase ($r$) is treated in the SSC scheme as an ellipsoidal inclusion embedded in an homogeneous equivalent medium whose behavior represents that of the polycrystal.
According to \citep{hill1965self,budiansky1965elastic}, tensor $\tenf{A}^{(r)}$ in phase ($r$) reads
\begin{equation}
\tenf{A}^{(r)}=\left \{
\tenf{I} + \tenf{S}_{\text{Esh}}:\left[\tenf{\widetilde{C}}\right ]^{-1}:
\left ( \tenf{C}^{(r)}-\tenf{\widetilde{C}} \right )
\right \}^{-1}\, ,
\label{eq:SSC_determination_A_r}
\end{equation}
with \tenf{I} the fourth order unit tensor.
The Eshelby tensor $\tenf{S}_{\text{Esh}}$ depends on $\tenf{\widetilde{C}}$ and on the aspect ratio $\ell / d $.
Here, the Eshelby tensor is calculated numerically as detailed in \citep{brenner2004mechanical}.
Equations (\ref{eq:C_eff_C_local_A}) and (\ref{eq:SSC_determination_A_r}) lead to an implicit equation for $\tenf{\widetilde{C}}$ that can be solved with a simple fixed-point method \citep{kroner1978self}.
Finally, from Eq.(\ref{eq:C_eff_C_local_A_r}), it can be observed that the sole knowledge of the mean (phase average) values $\tenf{A}^{(r)}$ is sufficient to estimate the overall behavior $\tenf{\widetilde{C}}$.
\textcolor{black}{
Therefore, computation of this mean-field SSC method is very fast, without having to know the complete field of $\tenf{A}(\vect{x})$. } 


\subsection{Mean-field generalized self-consistent scheme}
\label{sec:model_GSC}

\begin{figure}[!htbp]
\centering
\begin{tikzpicture}
\node[above=0cm] at (0,0) {\includegraphics[width=0.95\textwidth]{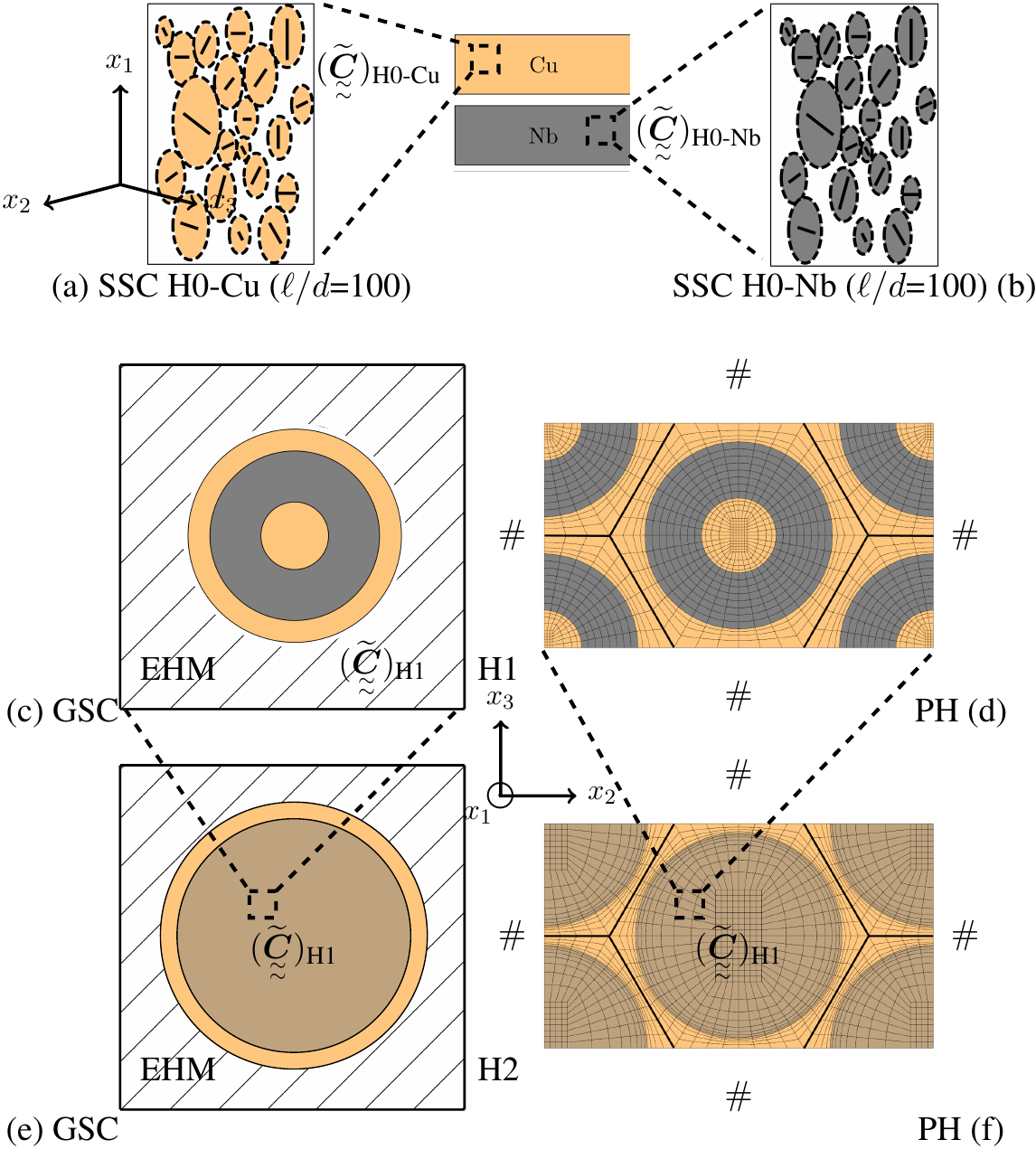}};
\end{tikzpicture}
\caption{\textcolor{black}{Multiscale modeling of effective elastic behaviors of Cu-Nb composite wires (Orange stands for Cu, gray for Nb, and brown for Cu-Nb composite).
(a,b) Cu and Nb polycrystals at the effective scale H0; the effective elastic tensor is obtained by the SSC scheme.
(c) At scale H1, using the GSC scheme, the Cu-Nb $3$-layered cylindrical inclusion is surrounded by an infinite Homogeneous Equivalent Medium; the Cu-Nb Composite Cylinders Assembly is assumed to exhibit a random fiber distribution in this model.
(d) Using the FEM PH model at scale H1, a periodic distribution is assumed ($\#$ denotes periodic boundary conditions).
(e) At scale H2, GSC scheme assumes a 2-layered cylindrical inclusion surrounded by a HEM.
(f) Using PH at scale H2, the distribution is periodic.
}}
\label{fig:Multi_Scale_Modeling_Zoom0-2_elastic}
\end{figure}

\textcolor{black}{
The ``($n+1$)-phase'' GSC scheme \citep{herve1995elastic} has been developed to estimate the overall effective elastic moduli of multi-coated fiber-reinforced composites with random fiber distribution; this model can also be used to estimate the overall effective elastic moduli of the studied materials because the moduli contrast between Cu and Nb is weak \citep{beicha2016effective}.} 
The GSC scheme was developed by considering at first the ``$n$-layered cylindrical inclusion problem'':
an n-layered cylindrical inclusion is embedded in an infinite matrix (i.e. phase $n+1$);
each phase is homogeneous, linearly elastic, transversely isotropic with the symmetry axis along the fiber direction $x_1$.
In addition, perfect bonding is assumed requiring the continuity of the stress vector and of the displacement field at the interfaces of different phases.
The above-defined specimen is subjected to several different remote boundary conditions in \cite{herve1995elastic} (so-called in-plane hydrostatic mode, normal tensile mode, in-plane transverse share mode and anti-plane longitudinal shear mode) aiming to derive the elastic strain and stress fields in each phase.
The infinite matrix, phase $n+1$, has been then replaced by an unknown HEM characterized by the effective elastic tensor $\widetilde{\tenf{C}}$.
This tensor is finally determined thanks to a self-consistent energy condition and to the previously solved ``$n$-layered cylindrical inclusion problem'' \citep{christensen1979solutions,herve1995elastic}.
In the present work, the effective elastic behavior of the studied Cu-Nb Composite Cylinders Assembly is computed by the GSC scheme  for H1 and H2 respectively as illustrated in Fig. \ref{fig:Multi_Scale_Modeling_Zoom0-2_elastic}(c,e).


\subsection{Full-field periodic models}
\label{sec:model_PH}

In addition to mean-field SSC and GSC schemes, a full-field FEM PH is proposed here to homogenize the effective elastic behavior of Cu-Nb wires at all scales (Fig. \ref{fig:overview_chart}).
An elementary volume $V$ made of heterogeneous material is considered for polycrystalline aggregates (scale H0) in Section \ref{sec:model_PH_poly} and for a specific architecture (scales H1-H3) in Section \ref{sec:model_PH_fiber}.
Periodic boundary conditions are prescribed at its boundary $\partial V$.
The displacement field $\vect{u}$ in $V$ takes the following form:
\begin{equation}
\vect{u}(\vect{x})=\bar{\tent{\varepsilon}}\cdotp\vect{x}+\vect{v}(\vect{x})~~~~\forall \vect{x} \in V
\label{eq:PH_periodic_conditions}
\end{equation}
where the fluctuation $\vect{v}$ is periodic, i.e. it takes the same values at two homologous points on opposite faces of $V$.
Furthermore, the traction vector $\tent{\sigma}\cdotp\vect{n}$ takes opposite values at two homologous points on opposite faces of $V$
($\vect{n}$ is the outwards normal vector to $\partial V$ at $\vect{x} \in V$).

Using the Voigt notation, stress and strain fields $\tent{\sigma}$ and $\tent{\varepsilon}$ are expressed as 6-dimensional vectors:
$\vect{\sigma}$=($\sigma_{11}$, $\sigma_{22}$, $\sigma_{33}$, $\sigma_{23}$, $\sigma_{13}$, $\sigma_{12}$)
and $\vect{\varepsilon}$=($\varepsilon_{11}$, $\varepsilon_{22}$, $\varepsilon_{33}$, $2\varepsilon_{23}$, $2\varepsilon_{13}$, $2\varepsilon_{12}$).
In order to determine the symmetric anisotropic tensor $\widetilde{\tenf{C}}$, six computations are necessary for each statistical realization of the volume element $V$ to find the 21 elastic coefficients \citep{kanit2003determination}.
Here, we impose successively six macroscopic normal and shear strain boundary conditions over $V$ as follows:
$\langle \vect{\varepsilon} \rangle$=$\vect{e_i}$ where $\vect{e_i}$ denotes the 6-dimensional unit vector, and $i$ varies from 1 to 6.
Then six homogenized stress tensors $\bar{\vect{\sigma}}$ can be determined by numerical homogenization leading to the effective elastic stiffness $\widetilde{\tenf{C}}$ by using Eq.(\ref{eq:local_effective_constitutive_relations}).

\subsubsection{PH adapted for polycrystalline aggregates}
\label{sec:model_PH_poly}

\begin{figure}[!htbp]
\centering
\begin{tikzpicture}
\node at (0,0) {\includegraphics[width=10.0cm]{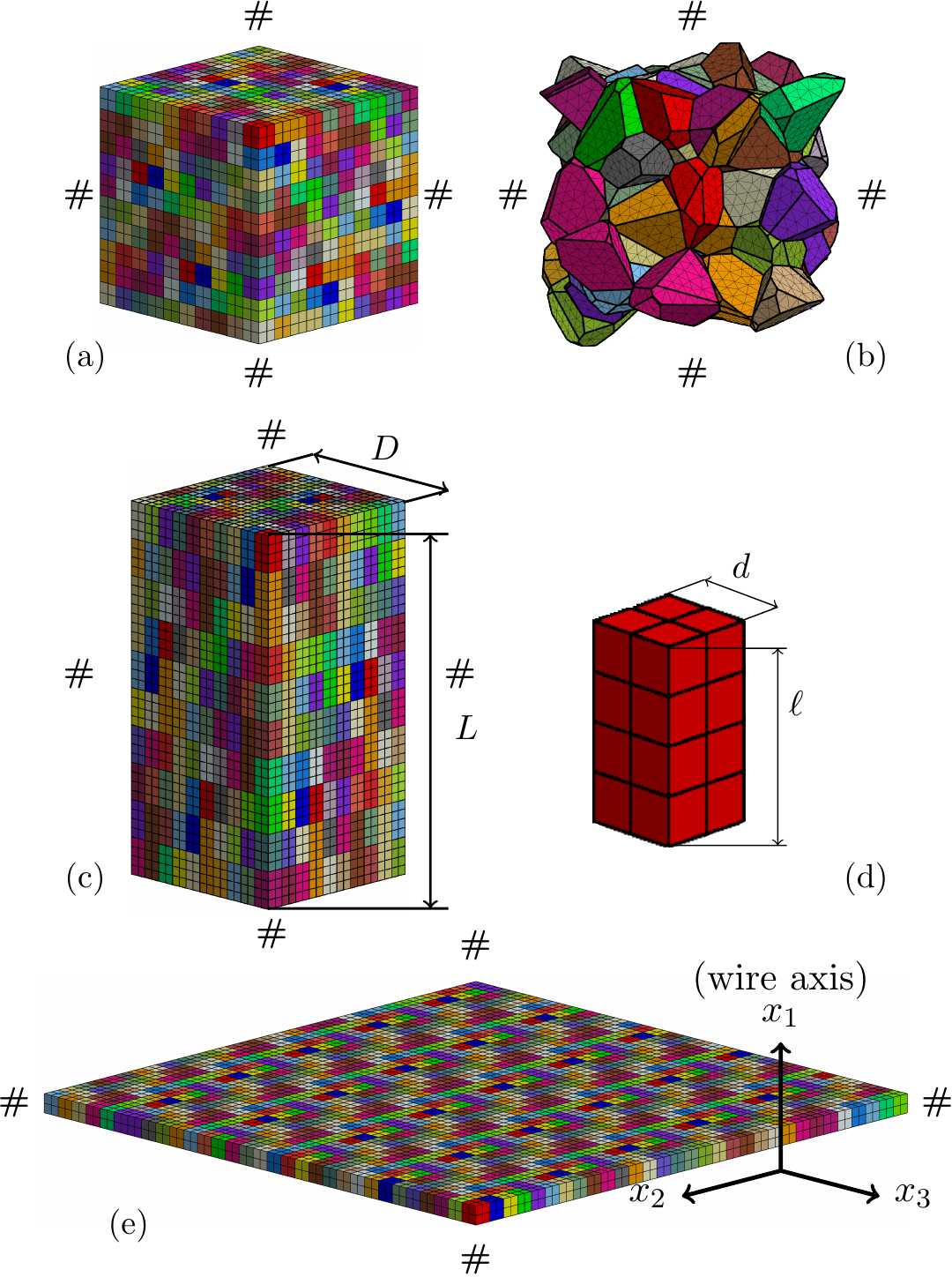}};
\end{tikzpicture}
\caption{Meshes of PH models for polycrystalline aggregates:
(a) parallelepipedic tessellation of (10 $\times$ 10 $\times$ 10) 1000 grains with the aspect ratio $\ell/d$=1;
(b) Vorono\"i tessellation of 100 grains with $\ell/d$=1.00$\pm$0.34;
(c) parallelepipedic tessellation of (10 $\times$ 10 $\times$ 10) 1000 grains with $L/D$=$\ell/d$=2,
$L$ and $D$ being defined as the length and width of the volume element respectively; tessellations of $L/D$=$\ell/d$=1,2,5,10 and 20 are considered in this work;
(d) Individual columnar grain for parallelepipedic tessellations in the case of $\ell/d$=2, elongated along the wire direction $x_1$;
(e) parallelepipedic tessellation of (31 $\times$ 31) 961 grains with $\ell/d \rightarrow \infty$.
Periodic boundary conditions, denoted \#, are considered.
}
\label{fig:PH_mesh}
\end{figure}

Unlike mean-field models in which the microstructure is described statistically, FEM PH can account for the real experimental microstructure at scale H0, and it provides the full-field stress/strain fields over $V$.
In the present work, PH is used to study the effect of a specific morphological/crystallographic texture and to compare the results with the ones obtained with mean-field methods.

As illustrated in Fig. \ref{fig:PH_mesh}(a), the polycrystalline aggregate is represented by a regular cubic grid made of parallelepipedic grains with the aspect ratio $\ell/d$=1.
Along each edge of this three-dimensional tessellation (finite element mesh using c3d20\footnote{c3d20: quadratic hexahedrons with 20 nodes per element}), 10 grains are considered, leading to $10 \times 10 \times 10$=1000 grains per mesh.
The used grain orientations (i.e. crystallographic texture) and single crystal properties of Cu and Nb are those given in Section \ref{sec:intro_textures} and Section \ref{sec:intro_notations}, and the discrete orientations are spatially randomly distributed among the grains of the parallelepipedic tessellation.
This elementary volume is subjected to periodic boundary conditions, Eq.(\ref{eq:PH_periodic_conditions}), in order to take advantage of a smaller RVE than the one with homogeneous boundary conditions \citep{kanit2003determination}.

Vorono\"i tessellation in Fig. \ref{fig:PH_mesh}(b) is also used to evaluate the impact of the over-simplified parallelepipedic grain shape in the previous problem.
This aggregate (finite element mesh using c3d4\footnote{c3d4: linear tetrahedrons with 4 nodes per element}) is subjected to periodic boundary conditions, and 100 Vorono\"i cells are distributed randomly in space.
The grain aspect ratio $\ell/d$ is statistically 1.00 with a 95\% confidence interval of $\pm\,0.34$.
\textcolor{black}{
Although more realistic than parallelepipedic grains, Vorono\"i tessellations still provide a smaller grain size distribution than real polycrystals \citep{lebensohn2005study,gerard2013modeling}.}

Due to the fabrication process, Cu and Nb grains are highly elongated along $x_1$.
In order to take this morphological texture into account, a series of parallelepipedic tessellations (c3d20) is also considered.
The case $L/D$=$\ell/d$=2, where $L$ and $D$ denote the mesh length and width respectively, is illustrated in Fig. \ref{fig:PH_mesh}(c).
Several meshes elongated along the wire direction $x_1$ are used in this work to consider various grain aspect ratios: $\ell/d$=2, 5, 10 and 20.
The case $\ell/d \rightarrow \infty$ is represented by a slice-shaped parallelepipedic tessellation composed of $31\times 31= 961$ grains subjected to periodic boundary conditions (Fig. \ref{fig:PH_mesh}(e)).
\textcolor{black}{Furthermore, the mesh density throughout this work has also been checked for ensuring adapted numerical accuracy.} 

When a single realization over the elementary volume $V$ is used, a relatively limited number of grain orientations and grain neighborhoods are accounted for.
This leads to a bias in the estimation of the effective properties as explained in \citep{kanit2003determination}.
The RVE \citep{hill1963elastic} must contain a sufficiently large number of heterogeneities (e.g., grains, inclusions, fibers ...) for the macroscopic properties to be independent on the boundary conditions applied to this volume.
\cite{kanit2003determination} proposed a statistical strategy to determine the RVE size.
Following this approach, we consider $N$ realizations of the microstructure in a volume with given size.
This volume size is then increased to investigate the asymptotic elastic behavior:
\begin{equation}
\overline{Z}=\displaystyle\frac{1}{N}\sum^{N}_{i=1}\widetilde{Z}_i,~~~~D_Z^2=\displaystyle\frac{1}{N}\sum^{N}_{i=1}(\widetilde{Z}_i-\overline{Z})^2
\label{eq:PH_RVE_mean_variance}
\end{equation}
where $\widetilde{Z}_i$ is an apparent elastic modulus obtained for one realization and $\overline{Z}$ is its mean value over $N$ realizations.
In addition, the variance $D_Z^2$ expresses the fluctuations of $\widetilde{Z}_i$.

The number of realizations $N$ is chosen so that the obtained mean value $\overline{Z}$ does not vary any longer up to a given precision when $N$ is increased.
This precision (i.e. relative error $\varepsilon_{\text{rela}}$) of the estimation of the effective property $\overline{Z}$  is related to the standard deviation $D_Z$ and the number of realizations $N$ by:
\begin{equation}
\varepsilon_{\text{rela}}=\displaystyle\frac{2D_Z}{\overline{Z}\sqrt{N}}\,.
\label{eq:PH_RVE_Relative_error}
\end{equation}

Conventionally, when $\varepsilon_{\text{rela}} \leq 1\%$, we suppose that the number of realizations $N$ is sufficiently large for being statistically representative of heterogeneous textures.
The RVE is then determined, and the overall effective elastic property is defined by the mean value over $N$ realizations, $\overline{Z}$.
In addition, the 95\% confidence interval is given by $[\overline{Z}-2D_Z, \overline{Z}+2D_Z]$.

For Cu polycrystals, the parallelepipedic tessellations require at least $N=3$ realizations for ensuring $\varepsilon_{\text{rela}} \leq 1\%$ for all of the five independent components of the elastic modulus.
\textcolor{black}{
In this work, the effective elastic tensor of $(\widetilde{\tenf{C}})_\text{H0-Cu}$ are determined by using $N=10$ random realizations, i.e. a total of 10000 (10$\times$1000) crystallographic orientations considered. } 
\textcolor{black}{
This has been done to reach a narrower confidence interval, which is a useful property for comparing results with those of the SSC scheme. }
We also choose 10 realizations for the slice-shaped tessellation with $\ell/d  \rightarrow \infty$.

On the other hand, each realization of the used Vorono\"i tessellation contains a smaller number of grains ($100$) than parallelepipedic tessellations.
For Cu polycrystal, it was found that $30$ random realizations are necessary to ensure statistical representativity of the results.
Therefore, the number of used Vorono\"i cells is 3000 ($30\times100$).
The same number of realization was used for Nb polycrystals.


\subsubsection{PH adapted for composite cylinders assemblies}
\label{sec:model_PH_fiber}

When using the analytic mean-field GSC scheme to determine the effective elastic moduli at scales H1 to H3, the long fibers are assumed to be distributed randomly.
In order to take into consideration the quasi-periodic fiber distribution observed experimentally (Fig. \ref{fig:Convention_multi_scales}) and to investigate the effect of this particular distribution, FEM PH will be also carried out for scales transitions H1 to H3 (Fig. \ref{fig:overview_chart}).
The section views of the unit cell of H1 and H2 are respectively indicated in Fig. \ref{fig:Multi_Scale_Modeling_Zoom0-2_elastic}(d) and Fig. \ref{fig:Multi_Scale_Modeling_Zoom0-2_elastic}(f).
Unit cells contain all information about the morphological RVE at the effective scales H1 and H2.
They are composed of two equivalent long fibers (1+4$\times$1/4 fibers) which are arranged in an hexagonal lattice, and they represent the (idealized) multi-scaled experimental microstructure of the Cu-Nb wires.
\textcolor{black}{Calculations require a 2D analysis with generalized plane strain conditions in order to allow for homogeneous strain in the third direction and also for out of plane shearing. For that purpose, a 3D mesh with one single element in the thickness (c3d20 elements) is used, together with suitable boundary conditions to keep flat upper and lower planes.}



\section{Homogenization results at scale H0}
\label{sec:H0}

In this section, the effective elastic behavior of Cu and Nb polycrystals, i.e. at scale H0, is considered.
Results of the various homogenization schemes presented above will be compared with each other.

As explained above, the anisotropic effective stiffness tensors $(\widetilde{\tenf{C}})_\text{H0-Cu}$ and $(\widetilde{\tenf{C}})_\text{H0-Nb}$ obtained by the SSC scheme and PH are found to be \textit{transversely isotropic} if we compute a statistically sufficiently large equivalent volume (i.e. RVE).
It should be noticed that, unlike the homogenized results of PH, the SSC ones display negligible scattering because of the very large number of grain orientations considered.

Five independent effective moduli were used to describe the overall anisotropic elastic behavior.
Results for two morphological textures ($\ell/d=1$, and highly elongated grains) are shown in Table \ref{tab:SSC_PH_ld1-inf_H0}.
In the case of equiaxed grains ($\ell/d=1$), it is remarkable that all the elastic moduli obtained for the three microstructures, i.e. SSC, parallelepipedic, and Vorono\"i tessellations, are in a perfect match, with a percentage difference
\footnote{percentage difference (in \%): the absolute difference between two values divided by their average.}
smaller than 4\%.
There are some visible differences of confidence intervals for PH homogenizations, the ones for parallelepipedic tessellations being about half those for Vorono\"i tessellations; this might be mainly caused by the computed number of grain orientations considered in these calculations.
It can also be observed that results of the SSC scheme fall within the confidence interval of PH, except for $\tilde\mu_{12}$ and $\tilde\mu_{23}$ for Nb polycrystals where slightly larger differences are found.

\begin{figure}[!htbp]
\centering
\begin{tikzpicture}
\node[above] at (0,0) {\includegraphics[width=0.75\textwidth]{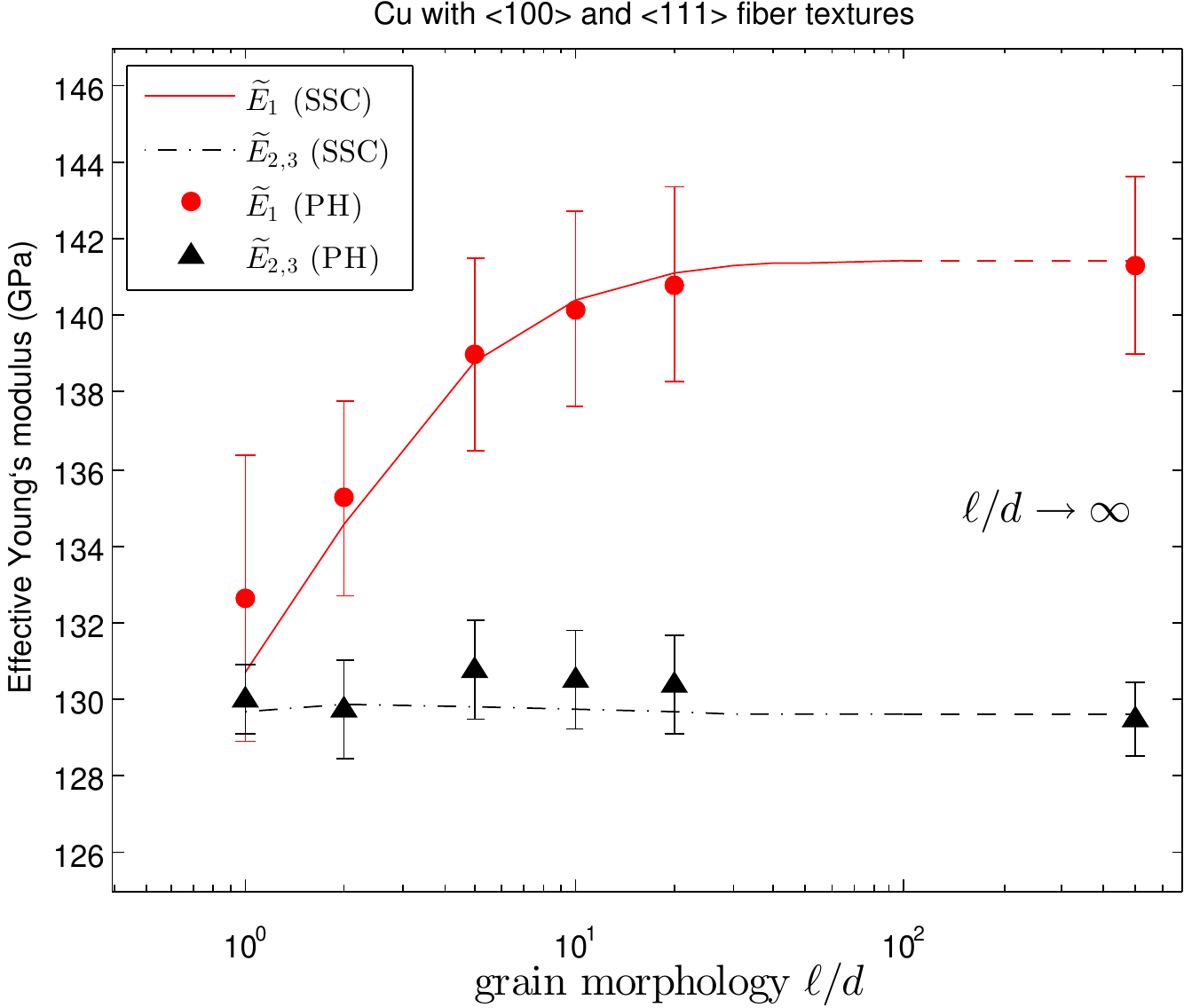}};
\node[below] at (0,0) {\includegraphics[width=0.75\textwidth]{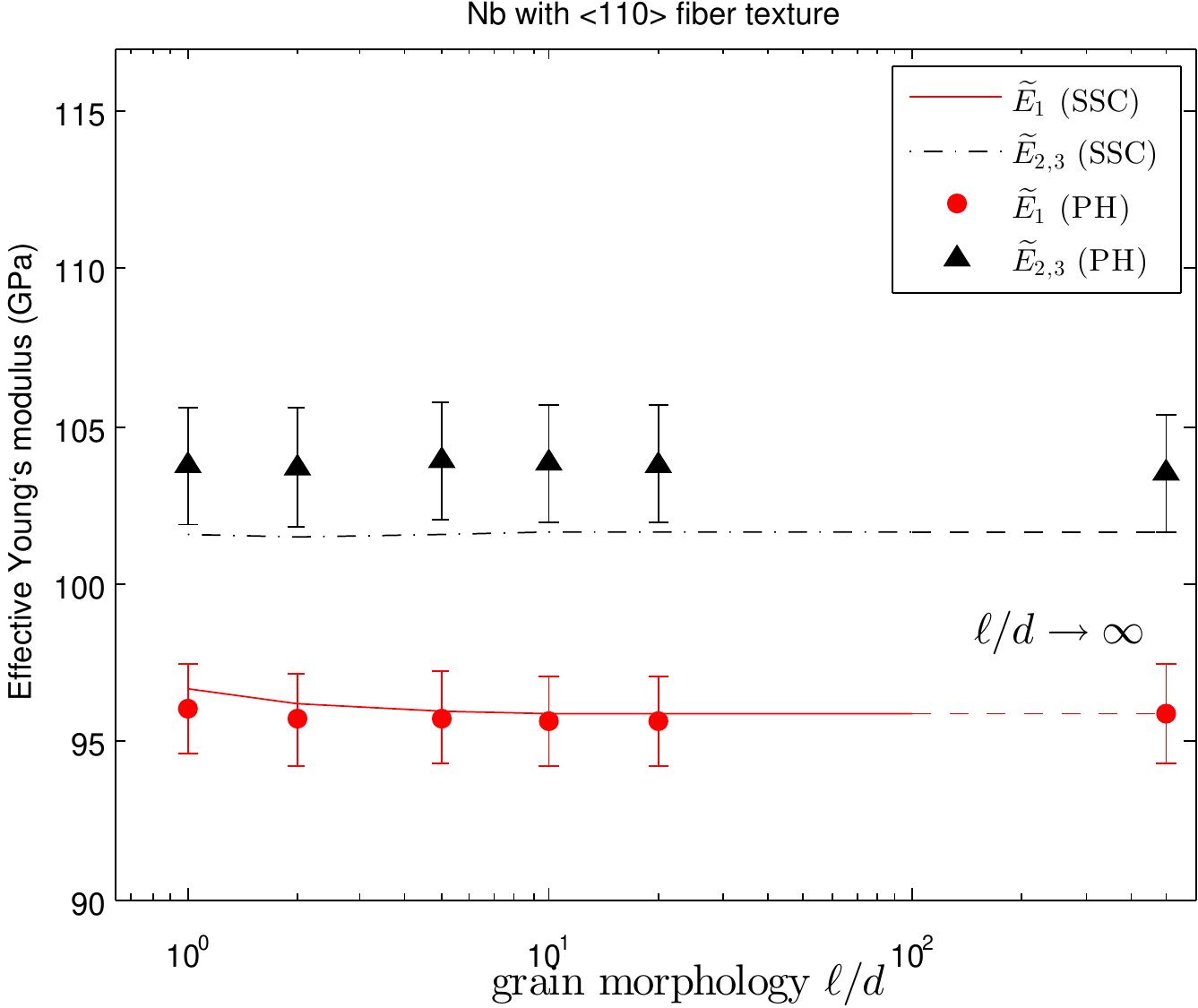}};
\end{tikzpicture}
\caption{Effective longitudinal and transverse Young's moduli ($\widetilde{E}_1$ and $\widetilde{E}_{2,3}$ respectively) in terms of the grain aspect ratio $\ell/d$.
The fiber textured Cu polycrystal and Nb polycrystal are homogenized by the SSC scheme and PH (using parallelepipedic tessellations).
}
\label{fig:SSC_PH_E-LT_ld}
\end{figure}

Figure \ref{fig:SSC_PH_E-LT_ld} further illustrates the SSC and PH (using parallelepipedic tessellations) predictions of longitudinal and transverse Young's moduli ($\widetilde{E}_1$ and $\widetilde{E}_{2,3}$) as a function of the grain aspect ratio $\ell/d$, for both Cu and Nb polycrystals.
Corresponding numerical values of the five independent effective moduli are provided in Table \ref{tab:SSC_PH_ld1-inf_H0} for $\ell/d =100$ (SSC scheme) and $\ell/d  \rightarrow \infty$ (PH).
Again, the agreement between all results is excellent.
It can however be noticed that a larger discrepancy is observed for the transverse modulus $\widetilde{E}_{2,3}$ obtained for Nb polycrystals.
Additional numerical tests will be performed in Section \ref{sec:discu_differe_SSC_PH_H0} to discuss the main factors that contribute to this (small) difference.
\begin{table}[!htbp]
\centering
\begin{tabular}{|c|c|c|c|c|c|}
\hline
Model & SSC & PH (para) & PH (Voro) & SSC & PH (para)\\
\hline
$\ell/d$ & 1 & 1 & 1.00$\pm$0.34 & 100 & $\infty$\\
\hline
\multicolumn{6}{|c|}{H0-Cu with a double fiber $\langle100\rangle$ and $\langle111\rangle$}\\
\hline
$\widetilde{E}_{1}$ ($\giga\pascal$) & 130.68 & 132.63$\pm$3.72 & 133.90$\pm$8.0 & 141.45 & 141.30$\pm$2.33\\
\hline
$\widetilde{\nu}_{12}$ & 0.340 & 0.338$\pm$0.006 & 0.336$\pm$0.012 & 0.327 & 0.327$\pm$0.004\\
\hline
$\widetilde{K}_{23}$ ($\giga\pascal$) & 152.37 & 152.79$\pm$1.08 & 152.95$\pm$2.39 & 153.88 & 153.91$\pm$0.82\\
\hline
$\widetilde{\mu}_{12}$ ($\giga\pascal$) & 47.65 & 48.30$\pm$0.80 & 49.40$\pm$1.80 & 46.43 & 46.30$\pm$0.52\\
\hline
$\widetilde{\mu}_{23}$ ($\giga\pascal$) & 48.20 & 48.11$\pm$0.61 & 49.17$\pm$1.14 & 46.86 & 46.80$\pm$0.64\\
\hline
\multicolumn{6}{|c|}{H0-Nb with a single fiber $\langle110\rangle$}\\
\hline
$\widetilde{E}_{1}$ ($\giga\pascal$) & 96.66 & 96.05$\pm$0.70 & 96.56$\pm$1.16 & 95.85 & 95.15$\pm$0.79\\
\hline
$\widetilde{\nu}_{12}$ & 0.408 & 0.410$\pm$0.004 & 0.409$\pm$0.005 & 0.408 & 0.411$\pm$0.003\\
\hline
$\widetilde{K}_{23}$ ($\giga\pascal$) & 185.75 & 184.54$\pm$1.22 & 184.78$\pm$2.28 & 185.65 & 184.48$\pm$1.28\\
\hline
$\widetilde{\mu}_{12}$ ($\giga\pascal$) & 39.60 & 38.36$\pm$0.54 & 38.69$\pm$1.23 & 39.48 & 38.32$\pm$0.62\\
\hline
$\widetilde{\mu}_{23}$ ($\giga\pascal$) & 36.88 & 38.27$\pm$0.54 & 38.44$\pm$1.30 & 37.04 & 38.26$\pm$0.62\\
\hline
\end{tabular}
\caption{Effective transversely isotropic moduli of Cu polycrystals and Nb polycrystals at scale H0 and their 95\% confidence intervals.
Experimental crystallographic textures and various morphological textures are considered for the SSC scheme and PH.
Both parallelepipedic (para) and Vorono\"i (Voro) tessellations are used for PH.}
\label{tab:SSC_PH_ld1-inf_H0}
\end{table}

Both SSC scheme and PH demonstrate remarkably that, for Cu polycrystals with the double $\langle100\rangle$-$\langle111\rangle$ fiber texture, aggregates with elongated grains display stiffer effective longitudinal Young's moduli than the ones with equiaxed grains.
For $\ell/d>20$, this tendency becomes saturated, and the elongation of grains along the wire direction $x_1$ has no more effect.
In contrast, for Nb polycrystals with the single $\langle110\rangle$ fiber texture, the effective Young moduli do not depend on the morphological textures.
The effects of morphological and crystallographic textures on the effective elastic behavior will be further discussed in Section \ref{sec:discu_influen_textures}.


\section{Effective behaviors at scales H1 to H3 of architectured Cu-Nb composites}
\label{sec:H1_to_H3}

\subsection{Results for scale H1}
\label{sec:H1_GSC_PH_SSC}

The SSC estimate is often advocated to be a good model for polycrystalline aggregates for two reasons:
(i) SSC scheme assumes a perfectly disordered mixture of grains which is similar to the real morphological textures in polycrystals;
(ii) SSC model can be applied to a statistically large heterogeneous volume composed of a very large number of grain orientations without costing much CPU time.
This has been confirmed by our results of the previous section.
Therefore, the homogenized anisotropic elastic tensors $(\widetilde{\tenf{C}})_{\text{H0-Cu}}$ and $(\widetilde{\tenf{C}})_{\text{H0-Nb}}$ for Cu polycrystals and Nb polycrystals determined by the SSC scheme will be taken here as local constitutive behaviors for the upper scale transitions (Fig. \ref{fig:overview_chart} and Fig. \ref{fig:Multi_Scale_Modeling_Zoom0-2_elastic}(a,b)).
Since grains in Cu-Nb wires are highly elongated with $\ell \gg d$, we consider within the SCC scheme a grain aspect ratio $\ell/d$=100, believed to be a good approximation of the columnar grains observed in the real microstructure (a larger aspect ratio does not change significantly the elastic properties).

We now proceed to the homogenization of the assembly of $85^1$ elementary continuum long fibers, i.e. at scale H1.
Both GSC and PH models are applied to the specific Composite Cylinders Assembly made of the co-cylindrical patterns with three layers: Cu-f/Nb-t/Cu-0, with properties at scales H0 provided by the SSC scheme as detailed above.
As presented previously in Fig. \ref{fig:Multi_Scale_Modeling_Zoom0-2_elastic}(c)(d), GSC scheme and PH assume that the Cu-Nb Composite Cylinders Assembly exhibits a random and a periodic fiber distribution, respectively.

Besides, for the sake of comparison, one can also use for scale H1 the simple SSC scheme, thus assuming a microstructure consisting of the sole random mixture of Cu and Nb grains, i.e. without consideration anymore of the specific architecture of the real specimen (Fig. \ref{fig:overview_chart}).
The volume fraction of Cu and Nb phases becomes 51.6\% and 48.4\%, respectively.
This corresponds to the normalized volume fraction of 10.5\%, 48.4\%, and 41.1\%  for Cu-f, Nb-t, and Cu-0, respectively (see Section \ref{sec:material_description}).

The transversely isotropic effective moduli $(\widetilde{\tenf{C}})_\text{H1}$ of the assembly of $85^1$ elementary long fibers (scale H1) are given in Table \ref{tab:SSC_GSC_PH_H1-H3}.
It is remarkable that the GSC scheme and PH provide very close results, and results of the SSC scheme are also in a perfect agreement.
The percentage difference between the prediction of these three models is less than only 2\%.

\begin{table}[htbp]
\centering
\begin{tabular}{|c|c|c|c|c|c|c|c|c|c|}
\hline
Scale & \multicolumn{3}{c|}{H1} & \multicolumn{3}{c|}{H2} & \multicolumn{3}{c|}{H3}\\
\hline
Model & SSC & GSC & PH & SSC & GSC & PH & SSC & GSC & PH \\
\hline
$\widetilde{E}_{1}$ ($\giga\pascal$) & 117.39 & 119.62 & 117.44 & 123.07 & 123.64 & 121.86 & 128.51 & 128.56 & 127.27 \\
\hline
$\widetilde{\nu}_{12}$ & 0.371 & 0.367 & 0.371 & 0.361 & 0.360 & 0.363 & 0.351 & 0.351 & 0.353 \\
\hline
$\widetilde{K}_{23}$ ($\giga\pascal$) & 169.56 & 168.07 & 169.56 & 165.67 & 165.33 & 166.51 & 162.06 & 162.06 & 162.89 \\
\hline
$\widetilde{\mu}_{12}$ ($\giga\pascal$) & 42.53 & 42.93 & 42.58 & 43.40 & 43.55 & 43.26 & 44.27 & 44.32 & 44.11\\
\hline
$\widetilde{\mu}_{23}$ ($\giga\pascal$) & 41.21 & 41.77 & 41.33 & 42.44 & 42.64 & 42.29 & 43.68 & 43.75 & 43.49\\
\hline
\end{tabular}
\caption{Effective transversely isotropic moduli of Cu-Nb wires at scales H1, H2 and H3
(i.e. Homogenization of the assembly of $85^1$, $85^2$ and $85^3$ elementary long fibers respectively),
obtained by mean-field SSC ($\ell/d$=100) and GSC schemes, and by full-field PH.}
\label{tab:SSC_GSC_PH_H1-H3}
\end{table}


\subsection{Iterative scale transition process up to scale H3}
\label{sec:up_to_H3}

At the effective scale H2, we suppose that the 85 continuum cylinders are composed of two layers: (1) the nano-composite Cu-Nb zones containing $85^1$ elementary long fibers; (2) the embedding matrix Cu-1.
In this work, an iterative process is proposed.
The effective tensor of the inner layer for scale H2, $(\tenf{C}^{\text{(1)}})_{\text{H2}}$, is given by the effective tensor $(\tenf{\widetilde{C}})_{\text{H1}}$ obtained for scale H1.
On the other hand, the effective behavior of the second layer for H2, $(\tenf{C}^{\text{(2)}})_{\text{H2}}$, is associated with the effective behavior of Cu polycrystals, $(\tenf{\widetilde{C}})_{\text{H0-Cu}}$.
The scale transition is then performed by GSC and PH approaches, leading to the effective tensor $(\widetilde{\tenf{C}})_\text{H2}$ for the assembly of $85^2$ elementary long fibers.

The same iterative process will be repeated up to scale H3 using GSC and PH approaches, allowing to estimate $(\widetilde{\tenf{C}})_\text{H3}$, the effective elasticity of 85 Cu-Nb composite zone of H2 embedded by Cu-2.
Moreover, as mentioned in Section \ref{sec:H1_GSC_PH_SSC}, regardless of the specific filament/nanotube microstructure, the SSC scheme can also be used to predict $(\widetilde{\tenf{C}})_\text{H2}$ and $(\widetilde{\tenf{C}})_\text{H3}$, considering the true volume fractions of Cu-Nb phases for scales H2 and H3, the crystallographic and morphological ($\ell/d$=100) texture, but discarding the material architecture.

The effective moduli of H2 and H3 are indicated in Table \ref{tab:SSC_GSC_PH_H1-H3}.
As before, SSC, GSC, and PH  homogenizations exhibit very close results at all the effective scales considered, the maximal percentage difference among them being as small as $\sim$1.5\%.
This result will receive further attention in Section \ref{sec:discu_archi_simpli}.


\section{Discussion}
\label{sec:discussions}

\subsection{SSC and PH predictions at scale H0}
\label{sec:discu_differe_SSC_PH_H0}

In this section, we investigate the factors that contribute to the deviation of the SSC scheme with respect to PH for Cu polycrystals and Nb polycrystals, i.e. considering scale H0.
Then in Section \ref{sec:discu_influen_textures}, the role of morphological and crystallographic textures on the effective elastic behavior will be discussed.

The effective elastic tensors $(\tenf{\widetilde{C}})_{\text{H0-Cu}}$ and $(\tenf{\widetilde{C}})_{\text{H0-Nb}}$ were determined in Section \ref{sec:H0} for the double $\langle100\rangle$-$\langle111\rangle$ fiber textured Cu polycrystal and the single $\langle110\rangle$ Nb polycrystal.
Both mean-field SSC scheme and full-field PH were applied, and an excellent agreement of model responses were found.
However, a larger percentage difference was found for $\widetilde{E}_{2,3}$ in the case of Nb polycrystals (see Fig. \ref{fig:SSC_PH_E-LT_ld}), a deviation that seems larger than for all other investigated moduli.
For a better understanding of the deviation between the SSC scheme and PH, additional numerical tests have been performed.

\begin{figure}[!htbp]
\centering
\begin{tikzpicture}
\node[above] at (0,0) {\includegraphics[width=0.75\textwidth]{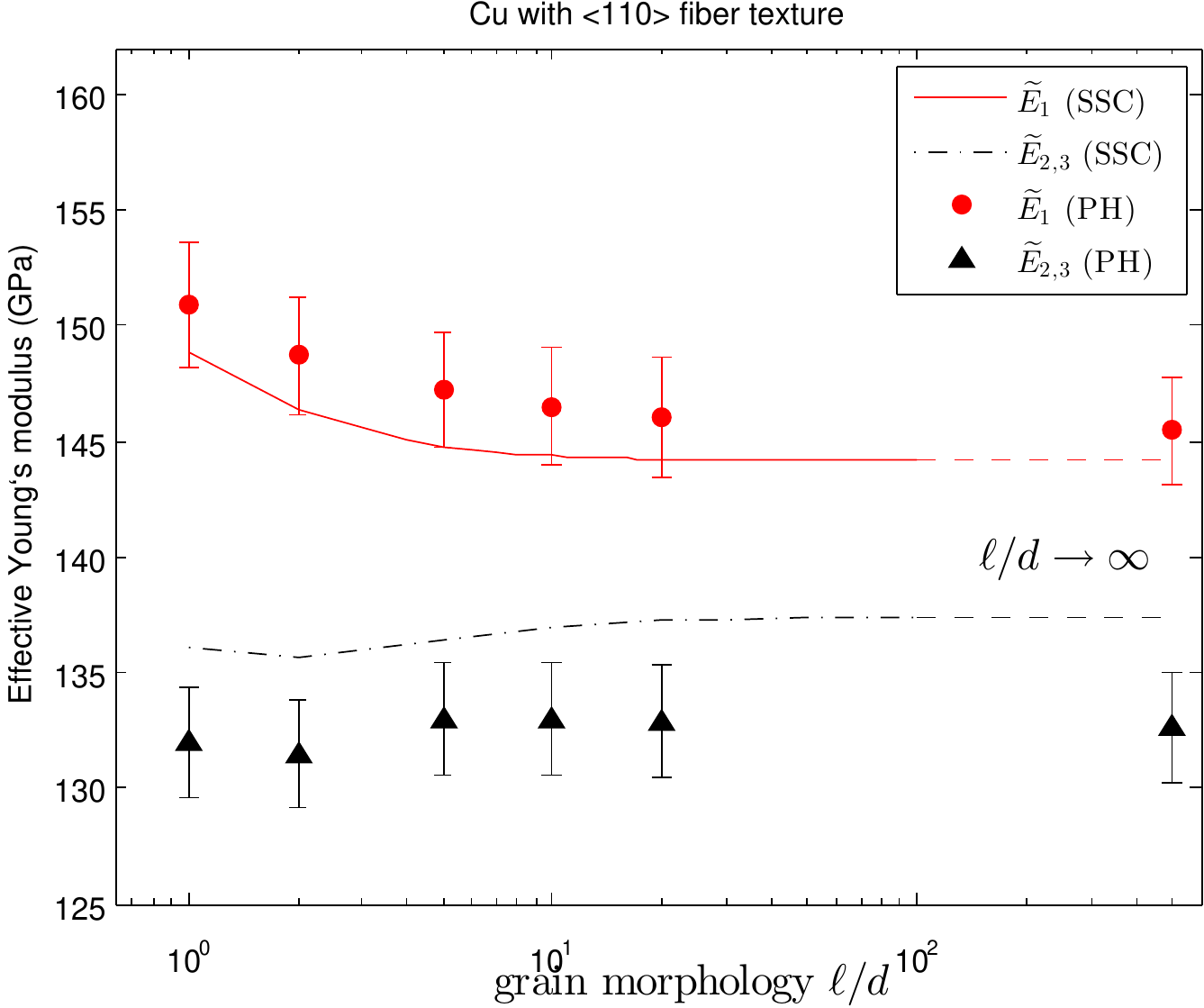}};
\node[below] at (0,0) {\includegraphics[width=0.75\textwidth]{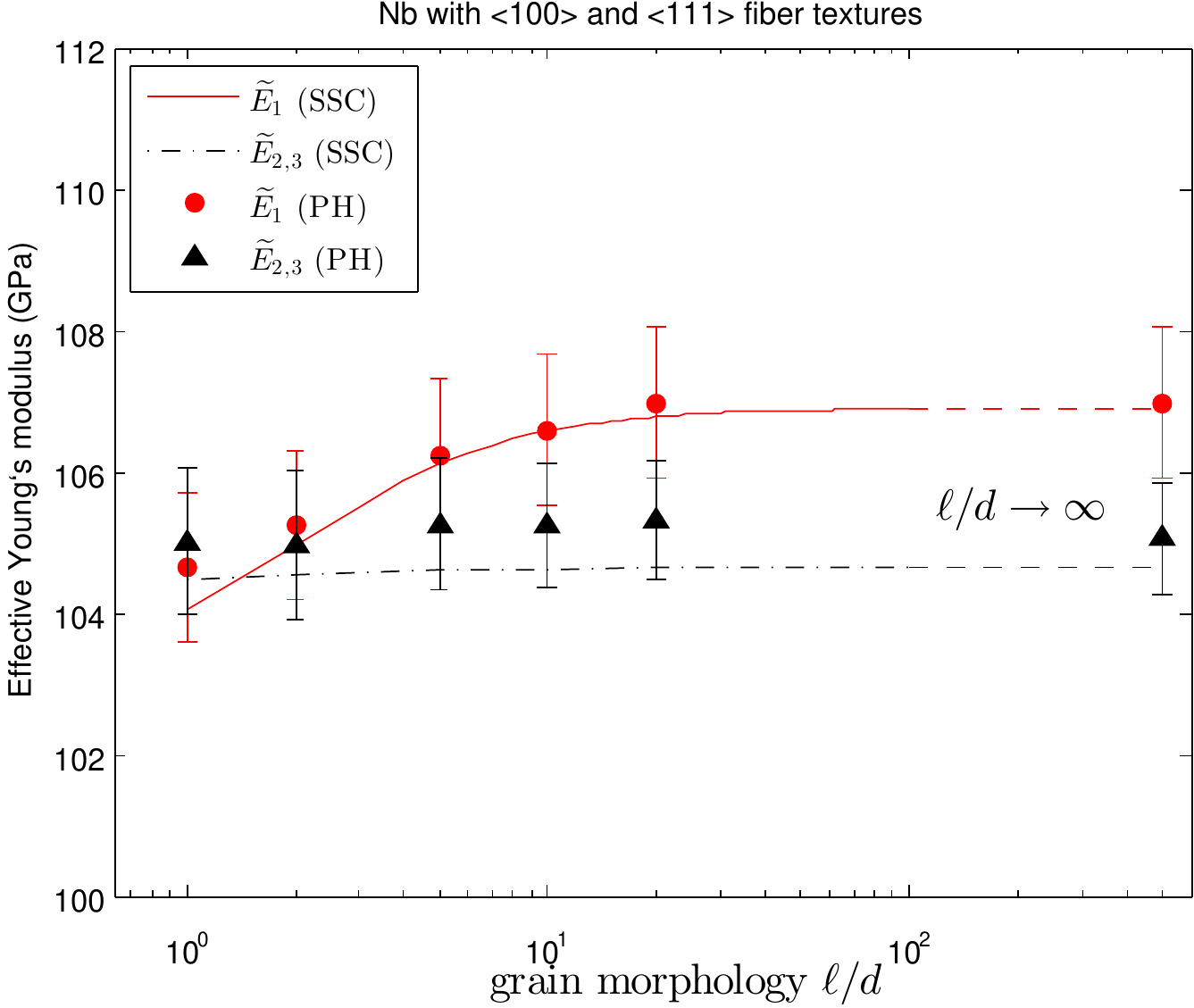}};
\end{tikzpicture}
\caption{Effective longitudinal and transverse Young's moduli ($\widetilde{E}_1$ and $\widetilde{E}_{2,3}$ respectively) in terms of the grain aspect ratio $\ell/d$, obtained by SSC and PH (using parallelepipedic tessellations).
Cu polycrystal and Nb polycrystal are textured by fictitious $\langle110\rangle$ and $\langle100\rangle$-$\langle111\rangle$ fibers respectively.
}
\label{fig:E_TL_ld_Cu_Nb_test}
\end{figure}

First of all, we exchanged the crystallographic textures : the experimental $\langle110\rangle$ fiber texture of Nb is taken as a fictitious crystallographic texture for Cu polycrystals.
Similarly, the double  $\langle100\rangle$-$\langle111\rangle$ fiber components of Cu are taken to build a fictitious polycrystal of Nb.
The predictions for the longitudinal and transverse Young's moduli ($\widetilde{E}_1$ and $\widetilde{E}_{2,3}$ respectively) are plotted in Fig. \ref{fig:E_TL_ld_Cu_Nb_test} as a function of the grain aspect ratio $\ell/d$ for these fictitious textures.

As before, it is observed that, despite the large difference of the Zener anisotropy factor $Z$ between Cu and Nb, SSC and PH models provide very similar results, for all the considered grain aspect ratio $\ell/d$ and crystallographic textures.
However, one sees that a larger discrepancy is now observed for the $\widetilde{E}_{2,3}$ modulus of Cu polycrystals, with the fictitious $\langle110\rangle$ fiber texture.
These differences mainly arise because these models do not take into account exactly the same grain topology: perfectly disordered mixture of grains for the SSC scheme and regular parallelepipedic grains of PH.
Grain size graduation is also infinite for the SSC scheme, whereas grains all have the same size for the parallelepipedic tessellations; grain size distribution of the Vorono\"i tessellation is narrow.
Combining Table \ref{tab:SSC_PH_ld1-inf_H0}, Fig. \ref{fig:SSC_PH_E-LT_ld} and Fig. \ref{fig:E_TL_ld_Cu_Nb_test}, it can be concluded that the effective properties of  sharp $\langle110\rangle$ fiber textures are more sensitive to microstructure details than double $\langle100\rangle$-$\langle111\rangle$ textures.


\subsection{Anisotropy induced by morphological and crystallographic textures}
\label{sec:discu_influen_textures}

In the preceding section, the SSC scheme has been shown to provide almost identical results than PH for various morphological and crystallographic textures.
Thanks to its high numerical efficiency, the SSC scheme will now be used to explore the role of micro-parameters, such as morphological and crystallographic textures.

\begin{figure}[!htbp]
\centering
\begin{tikzpicture}
\node[above] at (0,0) {\includegraphics[width=0.75\textwidth]{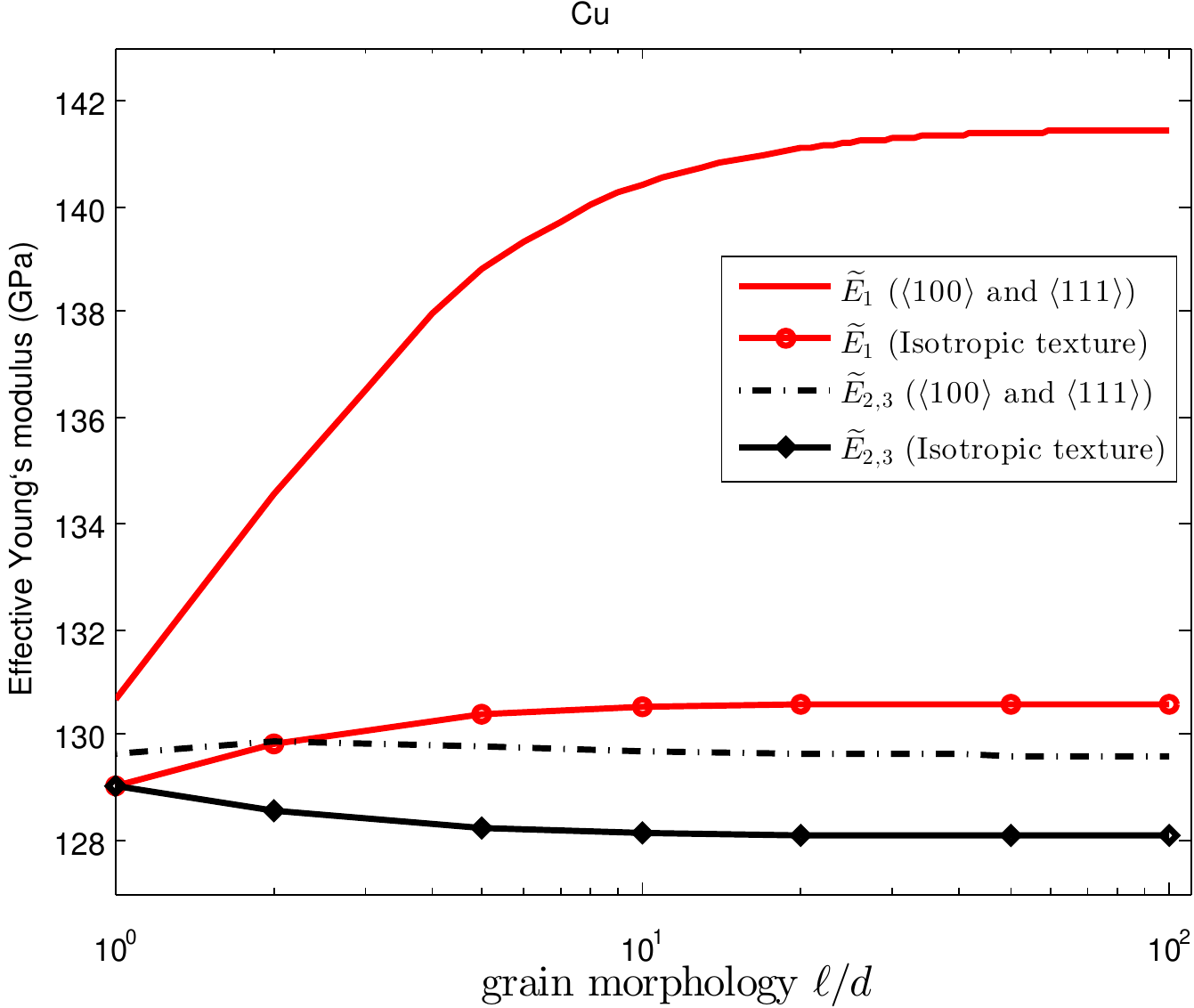}};
\node[below] at (0,0) {\includegraphics[width=0.75\textwidth]{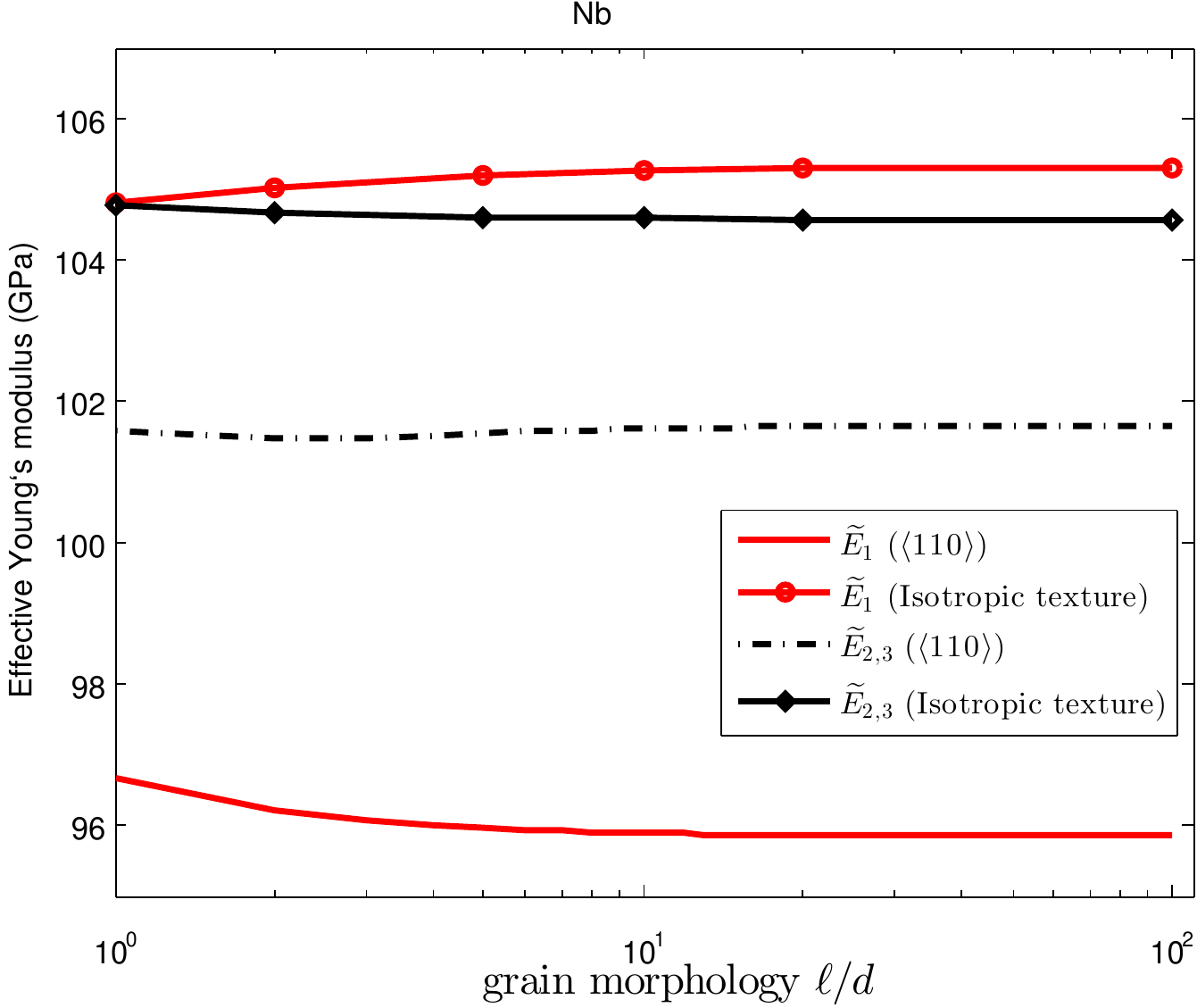}};
\end{tikzpicture}
\caption{Effective longitudinal and transverse Young's moduli ($\widetilde{E}_1$ and $\widetilde{E}_{2,3}$ respectively) in terms of the grain aspect ratio $\ell/d$ predicted by the SSC scheme.
Cu polycrystal and Nb polycrystal with experimental fiber textures are compared with the ones with a random (isotropic) texture.}
\label{fig:SSC_E_T-L_ld_isotropic_real}
\end{figure}

Fig. \ref{fig:SSC_E_T-L_ld_isotropic_real} shows predictions of the SSC scheme for the effective longitudinal and transverse Young's moduli ($\widetilde{E}_1$ and $\widetilde{E}_{2,3}$ respectively) as functions of the grain aspect ratio $\ell/d$, for Cu polycrystals and Nb polycrystals with a \emph{random} (i.e. isotropic) crystallographic texture.
Results are compared with the ones obtained for experimental fiber textures of Fig. \ref{fig:SSC_PH_E-LT_ld}.
For Nb polycrystals, it can be observed that grain morphology only has a small effect on the effective behavior, for both random and $\langle 110 \rangle$ crystallographic textures.
For Cu, the effect of grain morphology depends on the texture.
It has only a small influence for a random texture, but it affects significantly (by about $10\%$) $\widetilde{E}_1$ for the experimental $\langle 100 \rangle$-$\langle 111 \rangle$ texture.

\begin{figure}[!htbp]
\centering
\begin{tikzpicture}
\node[above] at (0,0) {\includegraphics[width=0.75\textwidth]{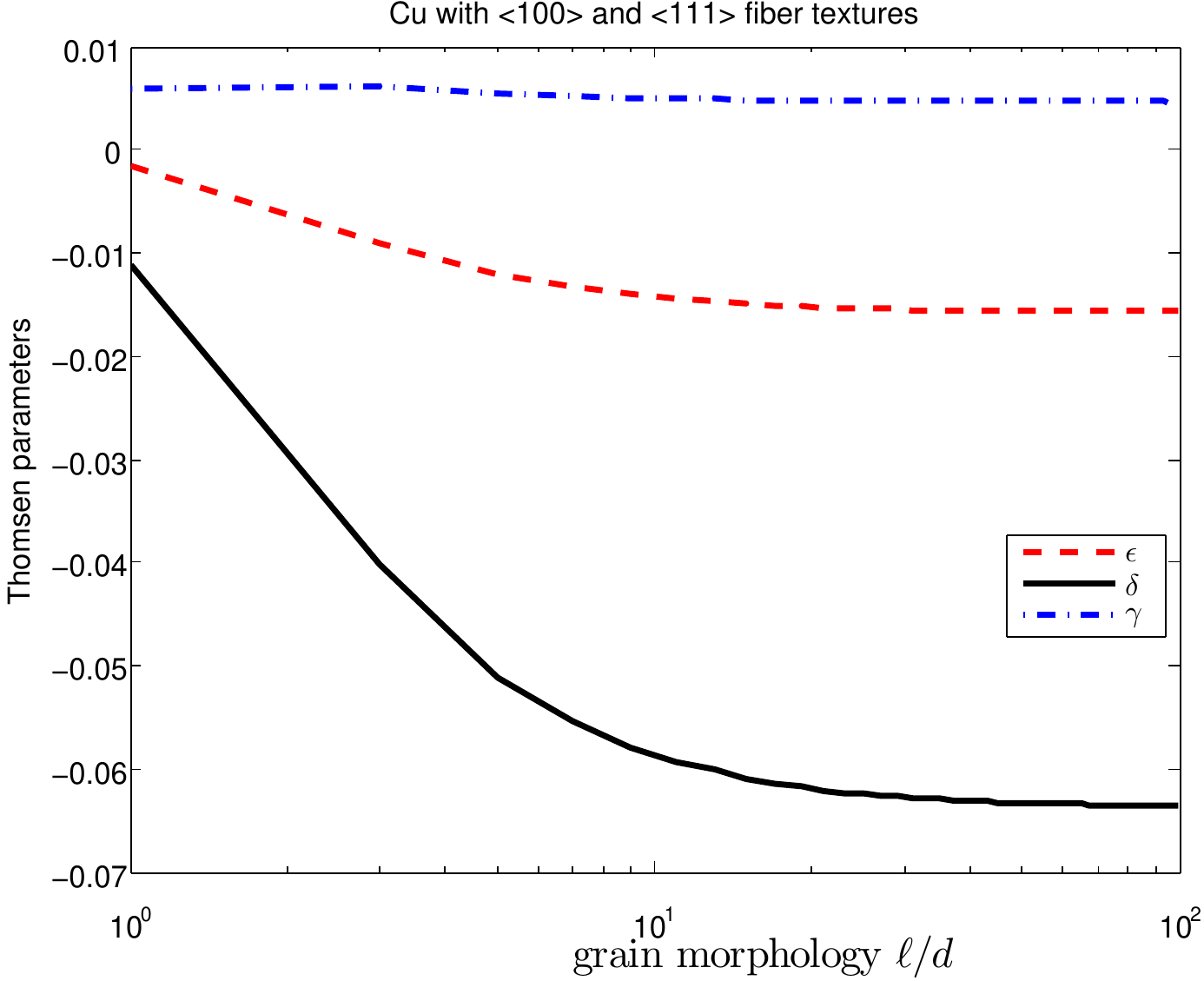}};
\node[below] at (0,0) {\includegraphics[width=0.75\textwidth]{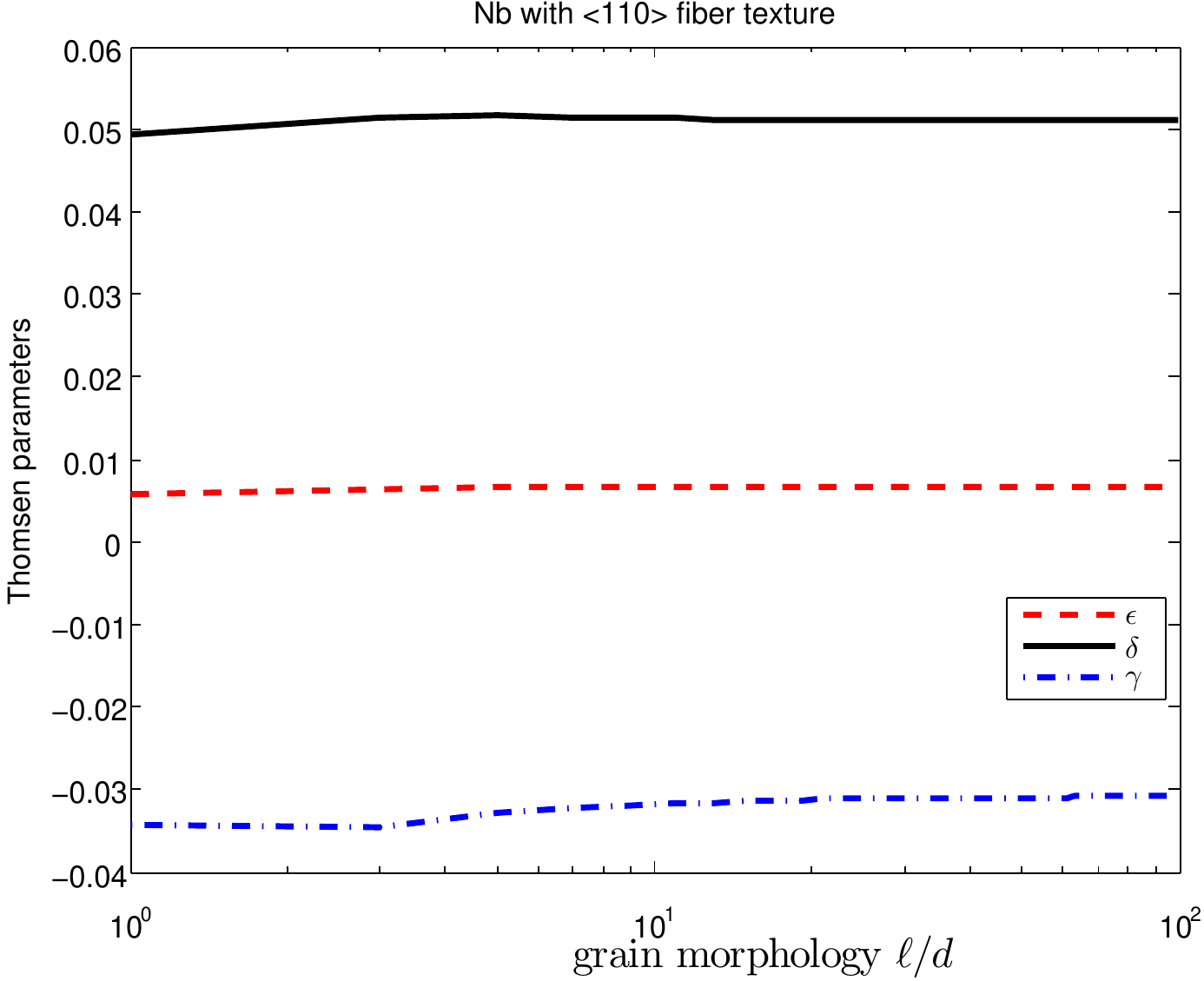}};
\end{tikzpicture}
\caption{Thomsen parameters in terms of the grain aspect ratio $\ell/d$, obtained with the SSC scheme.
Experimental fiber textures are used for Cu and Nb polycrystals.}
\label{fig:SSC_Thomasen_ld}
\end{figure}

To be more quantitative, Thomsen coefficients \citep{thomsen1986weak} are used for characterizing the transverse isotropy.
These dimensionless parameters are a combination of the components of the elastic stiffness matrix
\begin{equation}
\begin{aligned}
& \epsilon=\displaystyle{\frac{C_{33}-C_{11}}{2C_{11}}}\,, \\
& \delta=\displaystyle{\frac{(C_{13}+C_{66})^2-(C_{11}-C_{66})^2}{2C_{11}(C_{11}-C_{66})}}\,, \\
& \gamma=\displaystyle{\frac{C_{44}-C_{66}}{2C_{66}}}\,,
\end{aligned}
\label{eq:Thomsen_parameters}
\end{equation}
where index 1 indicates the symmetry axis ($x_1$).
For isotropic elasticity, the three Thomsen parameters are strictly equal to 0.
Conversely, the elastic mechanical behavior exhibits more anisotropy with larger absolute values of $\epsilon$, $\delta$ and $\gamma$.
Note also that the absolute value of these parameters is usually much less than 1.
Fig. \ref{fig:SSC_Thomasen_ld} illustrates these parameters obtained for $(\widetilde{\tenf{C}})_{\text{H0}}$ using the SSC scheme in terms of $\ell/d$ for the Cu and Nb polycrystals with experimental fiber textures.
It can be again observed that the behavior of the $\langle110\rangle$ fiber textured Nb is much less sensitive to grain morphology than the $\langle100\rangle$-$\langle111\rangle$ Cu.
For the latter one, the anisotropy is weak for equiaxed grain shape ($\ell/d$=1), and increases significantly with the grain aspect ratio.

\begin{figure}[!htbp]
\centering
\begin{tikzpicture}
\node[above] at (0,0) {\includegraphics[width=0.75\textwidth]{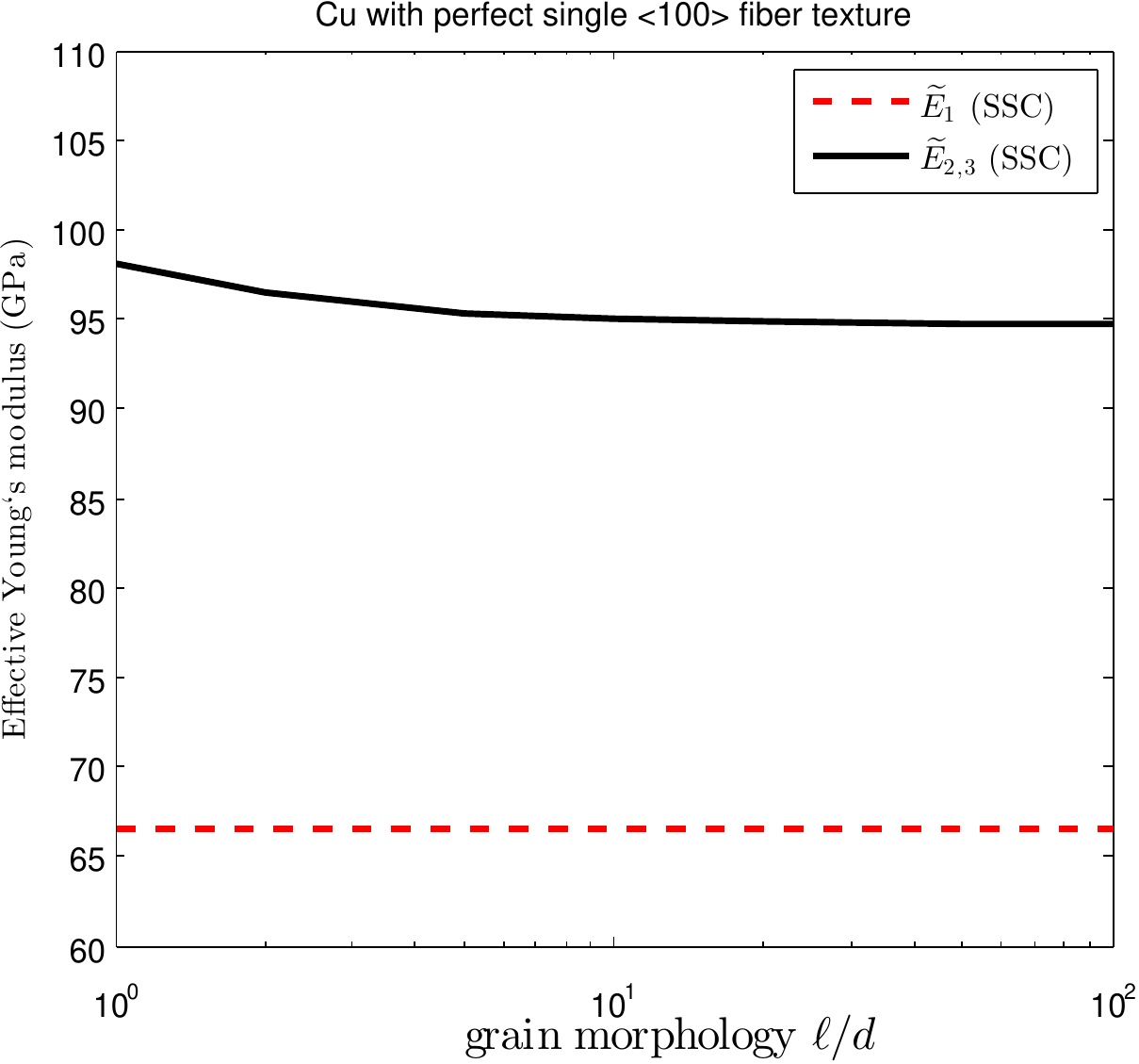}};
\node[below] at (0,0) {\includegraphics[width=0.75\textwidth]{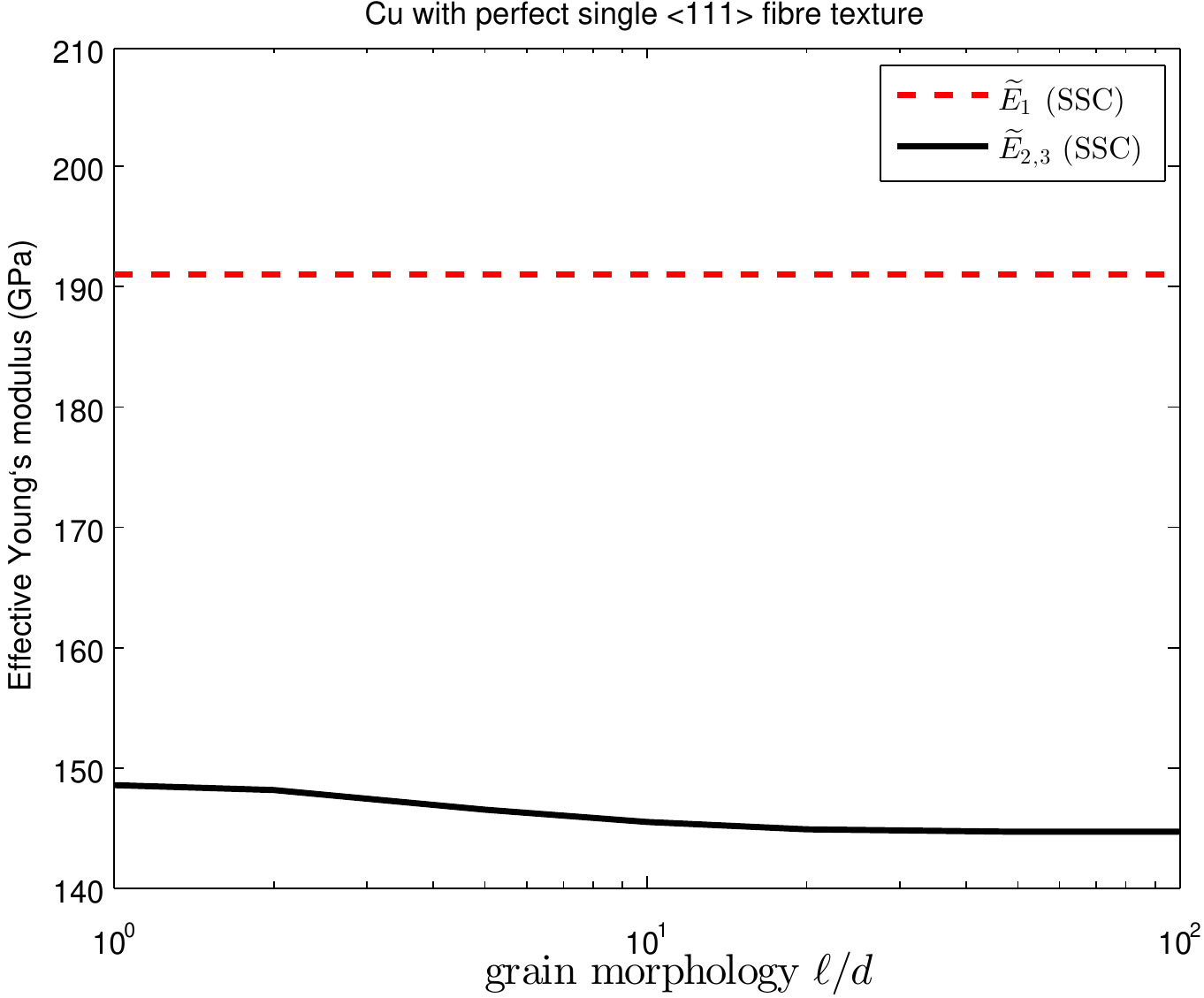}};
\end{tikzpicture}
\caption{Effective longitudinal and transverse Young's moduli ($\widetilde{E}_1$ and $\widetilde{E}_{2,3}$ respectively) of perfect single-component $\langle100\rangle$ and perfect single-component $\langle111\rangle$ fiber textured Cu by the SSC scheme in terms of $\ell/d$.}
\label{fig:E_TL_ld_Cu_100_111}
\end{figure}

The Cu polycrystals and Nb polycrystals with double $\langle100\rangle$-$\langle111\rangle$ fiber textures have been studied previously.
We now proceed to predict the effective elastic properties of \emph{perfect}\footnote{A texture component is \emph{perfect} when its spread FWHM vanishes, i.e. here all $\{100\}$ planes lie exactly parallel or perpendicular to $x_1$ axis.} $\langle100\rangle$ fiber textured aggregates and \emph{perfect} $\langle111\rangle$ ones separately for the propose of comparison.
An analytic solution for this homogenization problem has been derived by \cite{walpole1985evaluation}.
For Cu polycrystalline aggregates, one get
$\widetilde{E}_1$=$66.03\,\giga\pascal$, $\widetilde{E}_{2,3} \in \left [ 87.54, 105.79 \right ]\,\giga\pascal$ for a  $\langle100\rangle$ fiber texture, and
 $\widetilde{E}_1$=$191.49\,\giga\pascal$, $\widetilde{E}_{2,3} \in \left [ 129.82, 160.93 \right ]\,\giga\pascal$) for $\langle111\rangle$.
Fig. \ref{fig:E_TL_ld_Cu_100_111} illustrates the effective Young's moduli ($\widetilde{E}_1$ and $\widetilde{E}_{2,3}$) obtained by the SSC scheme for Cu polycrystals with various morphological textures.
These results, of course, are consistent with the analytical solution of Walpole.
It can be noted that $\widetilde{E}_1$ and $\widetilde{E}_{2,3}$ are significantly different for both texture components.
Moreover, it can be observed that, although $\widetilde{E}_1$ is insensitive to the grain aspect ratio for both individual $\langle100\rangle$ and $\langle111\rangle$ texture components, it become sensitive to the aspect ratio when the two components are mixed together (see Fig. \ref{fig:SSC_PH_E-LT_ld}).
In contrast $\widetilde{E}_{2,3}$ decreases with the aspect ratio for both individual texture components, but becomes rather insensitive to it when they are mixed together.

For sake of comparison, we have also computed the effective elastic parameters for Nb polycrystal with the perfect $\langle100\rangle$ fiber texture and the perfect $\langle111\rangle$ one:
$\widetilde{E}_1$=$145.14\,\giga\pascal$, $\widetilde{E}_{2,3}$=$117.69\,\giga\pascal$ for $\langle100\rangle$ and $\widetilde{E}_1$=$83.35\,\giga\pascal$, $\widetilde{E}_{2,3}$=$95.89\,\giga\pascal$ for $\langle111\rangle$.
It is found that these results
(which are in agreement with \cite{walpole1985evaluation} estimations:
$\widetilde{E}_1$=$145.48\,\giga\pascal$, $\widetilde{E}_{2,3} \in \left [ 113.62, 121.38 \right ]\,\giga\pascal$ for $\langle100\rangle$ and $\widetilde{E}_1$=$83.24\,\giga\pascal$, $\widetilde{E}_{2,3} \in \left [ 93.21, 98.28 \right ]\,\giga\pascal$ for $\langle111\rangle$)
are only slightly sensitive to the grain aspect ratio.
Note also that the difference in $\widetilde{E}_1$ and $\widetilde{E}_{2,3}$ for both texture components is less than for Cu, indicating a smaller effective anisotropy.

For Cu polycrystals, both effective Young's moduli ($\widetilde{E}_1$ and $\widetilde{E}_{2,3}$) and Thomsen parameters depend on the grain aspect ratio $\ell/d$ in the range of 1$\leqslant\ell/d <$20 irrespective of the fiber textures.
However, this dependency saturates with increasingly elongated grains along the wire direction $x_1$.
A similar feature has also been observed for Ni alloy directionally solidified polycrystalline aggregates in \cite{yaguchi2005accuracy}.
In contrast, the effect of the grain aspect ratio $\ell/d$ on the stiffness coefficients of Nb polycrystal is non-zero but rather small.

Concerning the effect of crystallographic texture, we have obtained different $\widetilde{E}_1$ values for different textures, for Cu polycrystals: $66.03\,\giga\pascal$ for perfect single fiber $\langle100\rangle$,
$\sim 146\,\giga\pascal$ for $\langle110\rangle$ (mean value between $\ell/d$=1 and $\ell/d \rightarrow \infty$), and $191.49\,\giga\pascal$ for perfect $\langle111\rangle$.
The percentage differences between the first two moduli and the last two ones are significant, i.e. 75\% and 26\% respectively.
For Nb polycrystals, $\widetilde{E}_1$ was found to be $145.14\,\giga\pascal$ for a perfect single $\langle100\rangle$ texture, $\sim 96\,\giga\pascal$ for single $\langle110\rangle$, and $83.35\,\giga\pascal$ for perfect single $\langle111\rangle$.
Thus, the percentage differences between the first two ones and the last two ones are less than for Cu (40\% and 15\% respectively), but still significant.

In summary, it was found that crystallographic textures play an important role in the effective elastic moduli of Cu and Nb polycrystals.
The combination of crystallographic and morphological texture effects is not straightforward.
For example, grains average morphology has strictly no effect on $\widetilde{E}_1$ for Cu with a $\langle100\rangle$ or a $\langle111\rangle$ perfect texture, but is responsible for a 10\% variation for the sharp $\langle100\rangle$-$\langle111\rangle$ experimental texture.
With the experimental textures, grain morphology has a much smaller effect on Nb that on Cu polycrystals.
These results highlight the necessity of accounting for the correct grain morphology and orientation when modelling the effective behavior at the different scales.
A comparison to experimental data is provided in the section \ref{sec:discu_S3_experimental_comparison}.


\subsection{Modeling Strategies for Cu-Nb wires}
\label{sec:discu_archi_simpli}

In section \ref{sec:H1_to_H3}, three homogenization models were applied to perform the scale transitions up to scale H3.
The SSC scheme assumes a random mixture of Cu and Nb phases;
the GSC scheme and PH both take into account the specific Composite Cylinders Assembly microstructure with random and periodic distribution, respectively.
As shown in Table \ref{tab:SSC_GSC_PH_H1-H3}, the three models provide very close results at H1, H2 and H3 scales, in spite of the very different approximations of geometry.

This good match is likely due to the relatively small elastic contrast between Cu and Nb elastic behavior.
For instance, the effective longitudinal Young's moduli ratio between Cu polycrystals and Nb polycrystals, $(\widetilde{E}_1)_{\text{H0-Cu}}/(\widetilde{E}_1)_{\text{H0-Nb}}$, is about 1.5 (Table \ref{tab:SSC_PH_ld1-inf_H0}).
Conversely, as shown by \cite{llorca2000elastic,beicha2016effective}, considerable deviations between SSC, GSC and PH models are obtained when the contrast is enlarged.
For instance in \cite{llorca2000elastic}, the Young's modulus obtained by the SSC scheme can be twice stiffer than the one obtained by GSC/PH model, for an epoxy matrix reinforced with steel spheres.
In this case, the Young's modulus of the spheres and of the matrix differ by a ratio of $E_\text{spheres}/E_\text{matrix}=60$.
For such a large contrast, the architectures and microstructures play a noticeable role on the effective elastic properties.

As a conclusion, the SSC scheme is demonstrated to be an efficient homogenization model for Cu-Nb wires, because of small elastic contrast between Cu phases and Nb phases.


\subsection{Structural problem S3 and experimental comparison}
\label{sec:discu_S3_experimental_comparison}

\begin{figure}[!htbp]
\centering
\includegraphics[width=8cm]{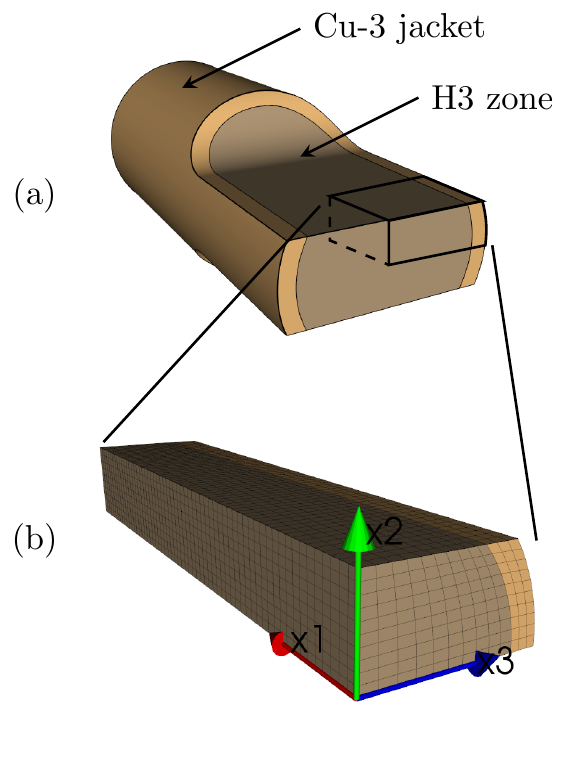}
\caption{Structural problem S3: (a) Schematic view of half of the S3 sample with a reduced section.
On such a sample geometry, multiple load-unload in-situ tensile tests were performed under X-rays by \cite{thilly2007evidence,thilly2009new}.
(b) 1/8th mesh of the reduced S3 sample.
}
\label{fig:Comparison_to_experimental_data}
\end{figure}

As illustrated in Fig. \ref{fig:Convention_multi_scales}(d) and detailed in Section \ref{sec:intro_scale_conven}, a real Cu-Nb wire can be seen as a single cylinder-shaped structure S3 at macro-scale.
In \cite{thilly2007evidence,thilly2009new}, the S3 sample was locally thinned to obtain a reduced gauge section below $0.14\,\milli\meter\squared$ over several millimeters long, as shown in Fig. \ref{fig:Comparison_to_experimental_data}(a).
This polishing allows X-rays to probe directly the nano-composite interior (i.e. so-called effective H3 zone) of the wire that would be otherwise screened by the external Cu-3 jacket.
It is worth noting that the polished H3 zone is now containing less than $85^3$ elementary fibers, but the material property, such as $(\widetilde{\tenf{C}})_\text{H3}$, is supposed not to be modified thanks to the quasi-periodic microstructure.
With this reduced S3 sample, multiple load-unload in-situ tensile tests were conducted \citep{thilly2007evidence}, and the experimental longitudinal Young's modulus $(\tilde E_{1})_\text{S3}$ was determined.
Obtained result ($131.2\,\giga\pascal$, uncertainty $\sim 5\%$) is given in Table \ref{tab:S3_experimental_comparison}.

The X-ray diffraction response of this reduced specimen has been investigated in \citep{thilly2007evidence,thilly2009new}.
By analysing the shift of $\{hkl\}$ Bragg peaks, X-ray diffraction provides information about the mean elastic strain within the diffracting volume.
The diffracting volume is constituted by all grains fulfilling Bragg conditions, i.e. exhibiting an $\{hkl\}$  plane perpendicular to the diffraction vector, i.e. perpendicular to the bisectrix of the incident and diffracted beams.
As (i) only a small proportion of grains fulfil Bragg conditions and (ii) diffracting grains exhibit a specific crystallographic direction, the diffraction volume does not provide a RVE.
\cite{thilly2007evidence,thilly2009new} took advantage of the strong crystallographic texture of the specimen for their analysis.
For Cu, the $\langle111\rangle$-$\langle100\rangle$ fibers lead to a significant proportion of  $\{220\}$ and $\{111\}$ planes respectively parallel and perpendicular to the wire axis $x_1$.
Performing diffraction on these lattice planes allows estimating the mean elastic strain, within the corresponding diffraction volumes,  along the specimen transverse and longitudinal directions, respectively.
For Nb, the strong $\langle110\rangle$ fibers leads to a significant proportion of  $\{220\}$ planes parallel or perpendicular to the wire, allowing estimating transverse and longitudinal mean elastic strains for the corresponding diffraction volumes.
Table \ref{tab:S3_experimental_comparison} reports results obtained for the case corresponding to a pure elastic response of the specimen, where transverse lattice strains on both Cu and Nb phases have been normalised by the applied effective tensile stress.
Uncertainty on $(\varepsilon_{\text{e}})_{\text{T}}/(\overline{\sigma}_{11})_{\text{S3}}$ is estimated as $\pm 5\%$ (fit uncertainty of  experimental data).

In addition, \cite{thilly2006plasticity} has performed neutron diffraction experiments on specimens similar to S3, i.e. on wires containing $55^4$ Nb/Cu-0 elementary long fibers, instead of $85^3$ Cu-f/Nb-t/Cu-0 fibers.
However, this specimen exhibits similar Cu/Nb volume fractions and textures to the previous ``co-cylindrical'' wires.
Diffraction on Cu$\{111\}$, Cu$\{220\}$, and Nb$\{110\}$ planes along different directions of the diffraction vector allows estimating the ratio $(\varepsilon_{\text{e}})_{\text{T}}/(\varepsilon_{\text{e}})_{\text{L}}$ between transverse and longitudinal strains, for both Cu and Nb phases, still for the corresponding diffraction volumes.
In \cite{thilly2009new}, uncertainties on the ``fine'' Cu channels (i.e. Cu-f, Cu-0 and Cu-1), the ``large'' Cu channels (i.e. Cu-2 and Cu-3), and the Nb-t, are estimated to be $\pm 25\%$, $\pm 13\%$ and $\pm 10\%$ respectively.
In our model, no distinction is made between fine and large Cu, treated as a single mechanical phase.
These uncertainties are reported in Table \ref{tab:S3_experimental_comparison}.

\begin{table}[htbp]
\centering
\begin{tabular}{|c|c|c|}
\hline
 & Experimental data & Modeling \\
\hline
$(\widetilde{E}_{1})_{\text{S3}}$ ($\giga\pascal$) & 131.2 $\pm 5\%$ & 130.6 \\
\hline
$(\varepsilon_{\text{e}}^{\text{\{220\}Cu}})_{\text{T}}/(\overline{\sigma}_{11})_{\text{S3}}$ ($\giga\pascal^{-1}$) & -2.2E-3 $\pm 5\%$ & -2.4E-3 \\
\hline
$(\varepsilon_{\text{e}}^{\text{\{110\}Nb}})_{\text{T}}/(\overline{\sigma}_{11})_{\text{S3}}$ ($\giga\pascal^{-1}$) & -4.0E-3 $\pm 5\%$ & -3.9E-3 \\
\hline
$(\varepsilon_{\text{e}}^{\text{\{220\}Cu}})_{\text{T}}/(\varepsilon_{\text{e}}^{\text{\{111\}Cu}})_{\text{L}}$ & -0.34  $\pm 13$-$25\%$ & -0.32 \\
\hline
$(\varepsilon_{\text{e}}^{\text{\{110\}Nb}})_{\text{T}}/(\varepsilon_{\text{e}}^{\text{\{110\}Nb}})_{\text{L}}$ & -0.62 $\pm 10\%$ & -0.51 \\
\hline
\end{tabular}
\caption{Comparison of experimental data with modeling results for macroscopic longitudinal stiffness $(\widetilde{E}_1)_\text{S3}$,
the ratio of transverse (i.e. along $x_2$-$x_3$ plane) elastic strain of diffracting planes of individual Cu/Nb phase to the macroscopic stress $(\varepsilon_{\text{e}})_{\text{T}}/(\overline{\sigma}_{11})_{\text{S3}}$
and the ratio of the transverse elastic strain to the longitudinal one
$(\varepsilon_{\text{e}})_{\text{T}}/(\varepsilon_{\text{e}})_{\text{L}}$.}
\label{tab:S3_experimental_comparison}
\end{table}

These data are now used to validate the multi-scale model presented above.
Fig. \ref{fig:Comparison_to_experimental_data}(b) indicates 1/8th of the thinned S3 sample (length $L$) with its mesh (c3d20).
For saving CPU time, symmetric boundary conditions are applied:
$U_1$ (displacement field along $x_1$) is fixed at the middle plane $x_1$=0;
$U_2$ is fixed at the bottom plane $x_2$=0
and $U_3$ is fixed at the left border plane $x_3$=0.
The elastic properties of the inner Cu-Nb composites zone of H3 and the external polycrystalline Cu-3 jacket are assigned to $(\widetilde{\tenf{C}})_\text{H3}$ (Table \ref{tab:SSC_GSC_PH_H1-H3}) and $(\widetilde{\tenf{C}})_\text{H0-Cu}$ (Table \ref{tab:SSC_PH_ld1-inf_H0} with $\ell/d$=100) respectively.
A normal tensile displacement $U_1$ is prescribed on the terminal section ($x$=$L$) of this S3 specimen.
From the computed axial force, longitudinal stiffness $(\widetilde{E}_1)_\text{S3}$ is obtained, in excellent agreement with the data (Table \ref{tab:S3_experimental_comparison}).

For a comparison with diffraction data, one need to (i) compute the mean stress acting on H3, (ii) compute the corresponding stress concentration within the grains, by means of the SSC scheme, and then (iii) isolate the diffraction volume and calculate the mean elastic strain along the direction of the different diffraction vectors.
Obtained results are indicated in Table \ref{tab:S3_experimental_comparison}, with again an excellent match to all available data.
From these experimental comparison, one can conclude that the proposed model not only predict the correct effective behavior, but also provide a good estimation of the stress and strain distributions within Cu and Nb grains; the mechanical scale transition is thus well captured.


\section{Conclusions}
\label{sec:conclusions}

The main conclusions of this work can be summarized as follows:
\begin{enumerate}
\item
\textit{Morphological texture effect}.
For the double $\sim$60\% $\langle111\rangle$ and $\sim$40\% $\langle100\rangle$ fiber textures, Cu polycrystal with elongated grains displays stiffer longitudinal Young's moduli than the one with spherical/square grains.
Furthermore, the parameters characterizing the anisotropy of the elastic properties are found to be higher for elongated copper grains.
In contrast, the effective elastic properties of the $\langle110\rangle$ Nb polycrystal only poorly depend on the morphological texture.
\item
\textit{Full field vs. mean field method}.
Regarding the effective behavior of the Cu polycrystals and Nb polycrystals separately, a good agreement is found between full-field FEM (PH) and SSC model up to only 3\% deviation for all considered crystallographic and morphological textures.
In addition, parallelepipedic tessellations provide very similar results as Vorono\"i ones.
\item
\textit{Morphological vs. crystallographic texture effect}.
Crystallographic texture plays an important role in the effective elastic moduli of Cu and Nb polycrystals.
Conversely, the morphological texture only slightly affects these effective moduli, except for some special crystallographic texture cases, such as the double $\langle100\rangle$-$\langle111\rangle$ fiber.
\item
\textit{Effective behavior at scale H1}.
The three homogenization schemes (i.e. SSC, GSC, and PH) provide very close results (differences smaller than 1.5\%) due to limited contrast of elastic properties in the Cu and Nb phases.
Therefore mean-field methods turn out to be more efficient than the full-field method PH for scale H1.
In spite of the absence of precise description of the microstructure, the SSC scheme delivers a good estimation of elastic properties.
\item
\textit{An iterative homogenization approach} is used to predict the elastic effective properties up to scale H3, and the elastic moduli estimated by the SSC scheme, the GSC scheme, and PH always almost coincide.
In the end, the real structure used in the in-situ X-rays experiments by \cite{thilly2007evidence,thilly2009new} is computed.
A very good agreement has been obtained between available experimental data and the numerical estimation for the effective longitudinal Young's modulus and for the elastic strain of
individual Cu/Nb reflection planes.
\end{enumerate}

Further work is in progress in the following directions:
The elastic homogenization methods of the present work will be extended to the elasto-plastic modeling of Cu-Nb wires, requiring full account of the different crystallographic textures from scales H0 to H3.
One of the challenge will be to account for the non-linear behavior at the different scale transitions.
\textcolor{black}{
Another challenge is to consider initial residual stresses induced by material processing.
Residual stresses have no influence on the elastic behavior investigated in this work, but play an important role in effective yield stress of Cu-Nb wires \citep{vidal2009plasticity}. }
Considering the grain size effect will also be necessary to predict the high stress levels reached in the plastic regime by these multi-scale materials.
Finally, the theoretical models will be used to optimize the microstructure parameters in the fabrication process of Cu-Nb wires.


\section*{Acknowledgment}
The authors gratefully acknowledge the support provided by ANR through the METAFORES ANR-12-BS09-0002 project.

\bibliographystyle{apalike}
\bibliography{Elasticity_IJSS}

\end{document}